\pdfoutput=1

\documentclass[prl,aps,twocolumn,showpacs,preprintnumbers,amssymb,nofootinbib,raggedbottom,10pt,longbibliography]{revtex4-2}

\usepackage{hyperref,graphicx,array,multirow,color,amsmath,mathrsfs,titlesec}
\titleformat{\section}{\centering\normalsize\normalfont\bf}{\thesection}{1em}{}

\usepackage[dvipsnames]{xcolor}
\usepackage[makeroom]{cancel}
\usepackage{amsmath,amssymb}
\usepackage[textsize=tiny]{todonotes}
\usepackage{pgfplots}
\pgfplotsset{compat=1.15}
\usepackage{tikz}
\usetikzlibrary{calc,3d,fillbetween}
\def\be{\begin{equation}}
\def\ee{\end{equation}}
\def\ba{\begin{eqnarray}}
\def\ea{\end{eqnarray}}
\newcommand{\bea}{\begin{eqnarray}}
\newcommand{\eea}{\end{eqnarray}}

\makeatletter\DeclareRobustCommand*{\bfseries}{\not@math@alphabet\bfseries\mathbf\fontseries\bfdefault\selectfont\boldmath}\makeatother

\begin{document}
\title{Antipodal Self-Duality for a Four-Particle Form Factor}

\author{Lance~J.~Dixon$^{1}$, {\"O}mer G{\"u}rdo{\u{g}}an$^{2}$, Yu-Ting Liu$^{1,3}$, Andrew~J.~McLeod$^{4,5}$ and Matthias Wilhelm$^{6}$}

\affiliation{$^1$ SLAC National Accelerator Laboratory,
Stanford University, Stanford, CA 94309, USA}

\affiliation{$^2$ School of Physics \& Astronomy, 
University of Southampton, Southampton, SO17 1BJ, UK}

\affiliation{$^3$ Kavli Institute for Theoretical Physics, UC Santa Barbara, Santa Barbara, CA 93106, USA}

\affiliation{$^4$ CERN, Theoretical Physics Department, 1211 Geneva 23, Switzerland}

\affiliation{$^5$ Mani L. Bhaumik Institute for Theoretical Physics, Department of Physics and Astronomy,
UCLA, Los Angeles, CA 90095, USA}

\affiliation{$^6$ Niels Bohr International Academy, Niels Bohr Institute,
  Blegdamsvej 17, 2100 Copenhagen \O{}, Denmark}

\preprint{CERN-TH-2022-190}
\preprint{SLAC-PUB-17711}

\begin{abstract}
We bootstrap the symbol of the maximal-helicity-violating four-particle form factor for the chiral part of the stress-tensor supermultiplet in planar $\mathcal{N}=4$ super-Yang-Mills theory at two loops. When minimally normalized, this symbol involves only 34 letters and obeys the extended Steinmann relations in all partially-overlapping three-particle momentum channels. 
In addition, the remainder function for this form factor exhibits an antipodal self-duality: It is invariant under the combined operation of the antipodal map defined on multiple polylogarithms---which reverses the order of the symbol letters---and a simple kinematic map. This self-duality holds on a four-dimensional parity-preserving kinematic hypersurface. It implies the antipodal duality recently noticed between the three-particle form factor and the six-particle amplitude in this theory.
\end{abstract}
\maketitle

\section{Introduction}\label{introduction_section}

Symmetries play a central role in modern formulations of fundamental physics, where they reflect simple facts about the world such as conservation laws and how different types of particles interact. However, sometimes new symmetries emerge in our theories whose physical implications are not immediately clear. These discoveries often lead to the development of more powerful mathematical techniques for making predictions, and have the potential to guide us to new physical principles. 

The planar limit of $\mathcal{N}=4$ super-Yang-Mills (SYM) theory has proven to be an especially rewarding place to look for (and exploit) novel symmetries in particle physics. Most famously, scattering amplitudes and form factors in this theory are dual to Wilson loops with lightlike edges~\cite{Alday:2007hr,Drummond:2007aua,Brandhuber:2007yx,Drummond:2007au,Alday:2008yw,Adamo:2011pv,Brandhuber:2010ad,Alday:2007he,Bern:2008ap,Drummond:2008aq,Maldacena:2010kp,Ben-Israel:2018ckc,Bianchi:2018rrj} and as such respect a dual conformal symmetry~\cite{Drummond:2006rz,Bern:2006ew,Bern:2007ct,Alday:2007hr,Drummond:2008vq}. These quantities also exhibit interesting number-theoretic symmetries~\cite{Caron-Huot:2019bsq}, as well as intriguing connections to cluster algebras, tropical fans, and positive geometries~\cite{Golden:2013xva,Golden:2014pua,Golden:2014xqa,Golden:2014xqf,Drummond:2017ssj,Bourjaily:2018aeq,Drummond:2018dfd,Golden:2018gtk,Drummond:2018caf,Golden:2019kks,Golden:2021ggj,Drummond:2019qjk,Drummond:2019cxm,Arkani-Hamed:2019rds,Henke:2019hve,Drummond:2020kqg,Mago:2020kmp,Chicherin:2020umh,Mago:2020nuv,Herderschee:2021dez,He:2021esx,Mago:2021luw,Henke:2021ity,Ren:2021ztg,Yang:2022gko}. While most of these special properties do not directly generalize to non-supersymmetric theories, their study has still led to significant improvements in our understanding of more general classes of amplitudes, form factors, and Feynman integrals (see for instance~\cite{Bourjaily:2019exo,Dixon:2020bbt,He:2020lcu,Chicherin:2020umh,He:2021eec,Abreu:2021smk,He:2022tph,Henn:2022ydo}). 

Much of this recent progress has been fueled by the in-depth study of amplitudes that evaluate to multiple polylogarithms (MPLs)~\cite{Chen,G91b,Goncharov:1998kja,Remiddi:1999ew,Borwein:1999js,Moch:2001zr}, which are endowed with a Hopf algebra structure~\cite{Gonch2,Goncharov:2010jf,Brown:2011ik,Brown1102.1312,Duhr:2011zq,Duhr:2012fh}. In particular, one part of the Hopf algebra of MPLs---the symbol map---greatly simplifies the study of what sequences of discontinuities appear in polylogarithmic functions. This fact has been leveraged to bootstrap certain amplitudes and form factors in planar $\mathcal{N}=4$ SYM theory to extremely high loop orders~\cite{Dixon:2011pw,Dixon:2011nj,Brandhuber:2012vm,Dixon:2013eka,Dixon:2014iba,Dixon:2014voa,Dixon:2015iva,Caron-Huot:2016owq,Dixon:2016nkn,Drummond:2018caf,Caron-Huot:2019vjl,Dixon:2020cnr,Dixon:2020bbt,Dixon:2022rse}. 

In an unexpected recent development, these bootstrap results have revealed a mysterious new duality between the maximally-helicity-violating (MHV) six-particle amplitude in parity-preserving kinematics, and the MHV three-particle form factor that involves a single insertion of the chiral stress tensor multiplet, which includes the Bogomol'nyi-Prasad-Sommerfield (BPS) operator $\text{tr}(\phi^2)$~\cite{Dixon:2021tdw}. Namely, these two quantities are related to each other by a map that appears in the Hopf algebra structure of MPLs: the antipode map.  At the level of the symbol, the antipode map simply reverses the order of discontinuities in MPLs. Thus, this duality can be loosely understood as the observation that the three-particle form factor and the MHV six-particle amplitude (in parity-preserving kinematics) encode exactly the same sequences of discontinuities, but in the opposite order---after a suitable map between the kinematic variables that describe the two processes.  At the moment, no physical argument is known for why this duality should hold, but it has been checked explicitly through at least seven loops~\cite{Dixon:2021tdw,Dixon:2022rse,ToAppearAmpEightLoop}. Interestingly, two-loop MHV amplitudes have also been shown to exhibit a different type of antipodal symmetry in parity-even kinematics, which is conjectured to hold to all particle multiplicity~\cite{Liu:2022vck}. 

In this paper, we describe a more general antipodal duality, which applies to four-particle form factors. Namely, the four-particle form factor is {\it self}-dual under the action of the antipode in a four-dimensional subspace of its kinematics. In fact, as we will show below, this new duality {\it implies} the duality between the three-particle form factor and the six-particle amplitude. The reason is that these quantities appear in the double and triple collinear limits of the four-particle form factor.

In order to substantiate this claim, we first bootstrap the symbol of the four-particle form factor at two loops. We do this by identifying the letters that appear in the Feynman integrals contributing to this form factor~\cite{Abreu:2020jxa,Abreu:2021smk,ToAppearAbreuEtal,samuel_private}. We then construct the space of integrable symbols for these letters. Amazingly, the form factor is uniquely identified in this space by just the first-entry condition, invariance under a standard set of discrete symmetries, and the strict (leading-power) double collinear limits. (This rigidity is reminiscent of the seven-particle amplitude, whose three-loop MHV symbol could be bootstrapped with only mild collinear information~\cite{Drummond:2014ffa}.) Three other limits---triple-collinear limits, the recently-computed limit of light-like operator momenta~\cite{Guo:2022qgv}, and the near-collinear limit---serve as cross-checks on our result. 

Multiple normalizations are needed to expose the different properties of this four-particle form factor.  In one normalization, the symbol of the form factor obeys the extended Steinmann relations (defined below) in all partially-overlapping three-particle momentum channels. In another normalization, this form factor is antipodally self-dual. We describe these normalizations, as well as the self-duality map, in more detail below. We provide the symbol in both normalizations in ancillary files, which also describe its various discrete symmetries and kinematic limits.

\section{The Bootstrap}
\label{sec:bootstrap}

 We begin our bootstrap by removing the infrared divergences and the MHV tree-level prefactor from the four-particle MHV form factor~\cite{Brandhuber:2011tv} $\mathcal{F}_4^{\text{MHV}}$,
 \begin{equation}
 \mathcal{F}^{\text{MHV}}_4 =  \mathcal{F}^{\text{min}}_4 \times F_4 \, ,
 \end{equation}
 where 
 \begin{equation}
 \mathcal{F}^{\text{min}}_4=
\mathcal{F}_4^{\text{MHV,\,tree}} \times \exp\biggl[ -\frac{g^2}{\epsilon^2} \sum_{i=1}^4 \left(\frac{\mu^2}{-s_{i,i+1}} \right)^\epsilon \biggr] \,.
 \end{equation}
Here, the 't Hooft coupling is $g^2 = N_c g_{\rm YM}^2/(16\pi^2)$. We have omitted contributions proportional to transcendental constants, because they vanish at the level of the symbol at which we are working.
Our main objective is to calculate $F_4$, which depends on eight dimensionless ratios:
 \begin{equation}
  u_{i}=\frac{(p_i+p_{i+1})^2}{q^2}\,,\quad v_{i}=\frac{(p_i+p_{i+1}+p_{i+2})^2}{q^2}\,,
 \end{equation}
where $i=1,2,3,4$ and $q=\sum_{i=1}^4 p_i$ is the (generically off-shell) momentum of the operator insertion. All indices should be understood to be mod 4. Due to momentum conservation and the masslessness of the four particles, $p_i^2 = 0$, these variables satisfy three constraints:
\begin{align}
-u_1 + u_3 + v_4 + v_1 &= 1  \, , \label{eq:uv_constraint_1} \\
-u_2 + u_4 + v_1 + v_2 &= 1 \, , \label{eq:uv_constraint_2} \\
-u_3 + u_1 + v_2 + v_3 &= 1 \, . \label{eq:uv_constraint_3}
\end{align}
Correspondingly, the four-particle form factor depends on five independent variables.

When expanded perturbatively in the coupling,
\be
F_4 = 1 + \sum_{L=1}^\infty g^{2L} F_4^{(L)} \,,
\label{eq:F4L}
\ee
the $L$-loop contribution $F_4^{(L)}$ is expected to be expressible as a linear combination of MPLs of weight $2L$ with rational coefficients.
The symbol of an MPL can be defined iteratively by its total differential~\cite{Goncharov:2010jf}:
  \begin{equation}
 d G = \sum_{x\in \mathcal{L}} G^x\, d \ln x \quad \Rightarrow \quad \mathcal{S}(G) = \sum_{x\in \mathcal{L}} \mathcal{S}(G^x\,) \otimes x \, ,
 \end{equation} 
 where $G^x$ are MPLs of one lower weight. The $d\ln$ arguments $x$ are referred to as symbol letters, while the total multiplicative span of the letters appearing in an MPL is referred to as its symbol alphabet, $\mathcal{L}$. For more background on MPLs and the symbol map, see for instance~\cite{Duhr:2014woa}.

The symbol of the one-loop form factor is~\cite{Brandhuber:2010ad}
\bea
\mathcal{S}\big(F_4^{(1)}\big) &=& 2 \, v_1 \otimes (1-v_1)
\ +\ \frac{u_1}{u_2 u_4} \otimes u_1  \nonumber\\
&&+\ \frac{u_1}{v_4 v_1} \otimes \frac{u_1-v_4 v_1}{u_1}
\ +\ \hbox{cyclic},
\label{eq:F41}
\eea
where the cyclic transformation maps $u_i\to u_{i+1}$ and $v_i\to v_{i+1}$.
We recall that the two-loop remainder function is related to the form factor itself by 
\begin{equation}
R_4^{(2)} = F_4^{(2)} - \frac{1}{2} \Bigl[ F_4^{(1)} \Bigr]^2 \,,
\label{eq:FRrel}
\end{equation}
and $R_4^{(2)}$ has smooth behavior in factorization limits \cite{Bern:2008ap,Dixon:2011pw,Brandhuber:2012vm}.

In order to bootstrap $F_4^{(2)}$ or $R_4^{(2)}$, we first assemble the alphabet of symbol letters that can appear in the four-particle form factor. Since the Feynman integrals that contribute to this form factor are all known~\cite{Abreu:2020jxa,Abreu:2021smk,ToAppearAbreuEtal}, this can be done easily.\footnote{While the nonplanar double pentagon Feynman integrals that contribute to this form factor have not yet been published~\cite{ToAppearAbreuEtal}, they do not give rise to letters beyond those that appear in the planar pentabox and nonplanar hexabox topologies~\cite{samuel_private}.} Altogether, the relevant Feynman integrals depend on 113 independent letters. Five different square roots appear in this alphabet, one of which involves the dihedrally-invariant argument 
\begin{equation}
\text{tr}_5 = 4 i \epsilon_{\alpha \beta \gamma \delta} p_1^\alpha p_2^\beta p_3^\gamma p_4^\delta \, , \label{eq:parity_root}
\end{equation}
and four of which are organized into pairs of two-orbits under the action of the dihedral group.\footnote{In the notation of~\cite{Abreu:2021smk}, these four additional roots correspond to two orientations of $\Delta_3$ and two orientations of $\Sigma_5$. Further orientations of these roots do not appear because they correspond to different planar orderings.}  The dihedral group $D_4$ is generated by the order-four cyclic transformation and a flip transformation, for example
$u_1 \leftrightarrow u_4$, $u_2 \leftrightarrow u_3$, $v_1 \leftrightarrow v_3$.
Spacetime parity acts by flipping the sign in front of $\text{tr}_5$, while leaving the sign in front of the other square roots intact.
 
We next construct the space of integrable weight-four symbols that draws upon this alphabet. We are only interested in symbols that have the correct branch cuts (satisfy the first-entry condition~\cite{Gaiotto:2011dt}), which in this context states that only the 8 letters $\{u_i,v_i\}$ can occur in the first entry. We also impose invariance under the dihedral symmetry group $D_4$.  There are 522 independent symbols $s_i$ satisfying these conditions, which we use to formulate our initial ansatz for the symbol of the remainder function $R_4^{(2)}$:
\begin{equation}
\mathcal{S}\big(R_4^{(2)}\big) = \sum_{i=1}^{522} c_i s_i \, , \label{eq:initial_ansatz}
\end{equation}
where the $c_i$ are undetermined rational coefficients. 

To fix the values of the coefficients in~\eqref{eq:initial_ansatz}, we first require that our ansatz is even under all elements of the algebraic Galois group, which flip the signs in front of each of the 5 square roots separately. (Note that the square-root signs are arbitrary conventions, on which the amplitude cannot depend. Note also that one of the elements is parity.) This imposes 148 independent conditions on the coefficients in our ansatz. Next, we require that our ansatz for $R_4^{(2)}$ reduces to the three-particle form-factor remainder $R_3^{(2)}$~\cite{Brandhuber:2012vm} when two of the external particles become collinear. 
In this limit, the 113-letter alphabet involves 25 spurious letters, in addition to the 6 letters describing the three-particle form factor $F_3$. Matching our ansatz onto the correct expression completely fixes
the remaining 374 coefficients, and thus uniquely determines the
symbol of $\smash{R_4^{(2)}}$, or equivalently of $\smash{F_4^{(2)}}$.
The sparse systems of linear equations that encode these constraints
can be solved efficiently over finite fields using the {\sc SpaSM}
software library~\cite{spasm}.
The numbers of free parameters at each stage in
the calculation are collected in Table~\ref{tab:constraints}.

Although we started with an initial ansatz of over one hundred letters, our result for the symbol of $F_4^{(2)}$ only involves 34 letters (notably, all five square roots still appear).  As expected, it also obeys the Steinmann relations~\cite{Steinmann,Steinmann2}, which forbid sequential discontinuities in partially-overlapping momentum channels; in the case of massless scattering amplitudes, these relations apply when both channels contain at least three particles~\cite{Caron-Huot:2016owq}. However, while the Steinmann relations only put constraints on the first two entries of the symbol, many amplitudes have been found to obey an {\it extended} set of Steinmann relations, in which the same constraints hold for all adjacent entries in the symbol~\cite{Caron-Huot:2018dsv,Caron-Huot:2019vjl,Caron-Huot:2019bsq,He:2021mme}. In processes involving one massive and four massless external legs, the (extended) Steinmann relations forbid the letter $v_i$ from appearing next to $v_j$ in the symbol when $j\neq i$.
Notably, the two-loop master integrals that contribute to $F_4^{(2)}$ only obey the Steinmann relations---not the extended Steinmann relations---between $v_i$ and $v_{i+2}$~\cite{Abreu:2021smk}.  However, the extended Steinmann relations are obeyed in \emph{all} channels by $\smash{F_4^{(2)}}$. Similarly, higher-point amplitudes in this theory obey all extended Steinmann relations when normalized minimally (although amplitudes do not respect dual conformal invariance in this normalization)~\cite{Golden:2019kks,Caron-Huot:2019vjl}.

\begin{table}[t]
 \begin{tabular}{|l|c|}
\hline
Constraints   & Parameters\rule[-3pt]{0pt}{13pt} \\\hline\hline

first entry, integrability, $D_4$ invariance   & 522\rule[-3pt]{0pt}{13pt} \\\hline
Galois symmetry       &          374\rule[-3pt]{0pt}{13pt} \\\hline
strict double collinear limit $\to R_3^{(2)}$ & 0\rule[-3pt]{0pt}{13pt} \\\hline
strict triple collinear limit $\to \hat{R}_6^{(2)}$  & 0\rule[-3pt]{0pt}{13pt}\\\hline
light-like limit       &      0\rule[-3pt]{0pt}{13pt}\\\hline
FFOPE        &     0\rule[-3pt]{0pt}{13pt}\\\hline
 \end{tabular}
\caption{Number of free parameters that remain after imposing each constraint
in the bootstrap procedure, starting with the 113-letter alphabet. 
The limits are taken at leading power, except for the FFOPE.}
\label{tab:constraints}
\end{table}
 
\section{Special Kinematic Limits}
\label{sec:kinematiclimits}

We can check our results in a number of different kinematic
limits. First we consider the light-like limit, where the
operator momentum $q^2 \to 0$. In this limit, the 34 letters that
appear in the symbol of $R_4^{(2)}$ reduce to 13
independent multiplicative combinations. Nine of these combinations match the light-like letters reported in~\cite{Guo:2022qgv}, while four are spurious and must drop out of $R_4$. We have confirmed that our $\smash{R_4^{(2)}}$ symbol correctly
reproduces the symbol of the light-like form factor remainder reported in~\cite{Guo:2022qgv}.
 
Although our bootstrap procedure made use of information about the strict, or leading power, collinear limit, we can still make nontrivial predictions for the subleading powers in the expansion of $\smash{F_4^{(2)}}$ around this limit. Such terms are predicted by the recently-developed Form Factor Operator Product Expansion (FFOPE)~\cite{Sever:2020jjx,Sever:2021nsq,Sever:2021xga}. To carry out this cross-check, we rewrite our kinematic variables in terms of the OPE variables $T$, $T_2$, $S$, $S_2$, and $f_2$~\cite{Basso:2013vsa,Sever:2020jjx}:
\begin{align}
  u_1&= \frac{T^2 T_2^2}{\left(T^2+1\right) \left(S^2+T^2+T_2^2+1\right)} \, ,
\nonumber\\
  u_2&= \biggl[1 {+} T^2 {+} \frac{S^2 [S_2 T_2 (1{+} f_2^2 ){+} f_2 (1 {+} S_2^2 {+} T^2 {+} T_2^2)]}{f_2 S_2^2}\biggr]^{-1} \!\! , \nonumber \\
  u_3&= \frac{S^2}{\left(T^2+1\right) \left(S^2+T^2+T_2^2+1\right)} \, ,
\label{eq:ope_parametrization} \\  
  u_4&= \frac{S^2 T^2}{S_2^2} u_2 \, ,  \nonumber \\
  v_1&= \frac{T_2^2+1}{S^2+T^2+T_2^2+1}\, , \nonumber
 \end{align}
while $v_2$, $v_3$, and $v_4$ are fixed by the relations~\eqref{eq:uv_constraint_1}--\eqref{eq:uv_constraint_3}. The near-collinear limit in these variables corresponds to an expansion around small values of $T_2$, which we have computed by expanding our symbol to $\mathcal{O}(T^2 T_2^2)$. We checked this expansion against the predictions made by the FFOPE (using the procedure explained in~\cite{Sever:2021nsq}) for the $T^2T_2\ln(T)$,  $T^2T_2\ln(T_2)$,  $T^2T_2$, and $T^2T_2^2\ln(T_2)$ contributions.\footnote{We note that the contributions at $\mathcal{O}(T^0)$ simply reproduce the OPE expansion of $\hat{R}_6$, while the contributions at $\mathcal{O}(T_2^0)$ reproduce the OPE expansion of $R_3$.}
Each of these checks was carried out as a series in $S$ and $S_2$ and to all available powers of $f_2$.

Finally, it can be seen from the OPE~\cite{Alday:2010ku,Basso:2013vsa,Basso:2013aha,Basso:2014koa,Basso:2014jfa,Basso:2014nra,Belitsky:2014sla,Belitsky:2014lta,Basso:2014hfa,Belitsky:2015efa,Basso:2015rta,Basso:2015uxa,Belitsky:2016vyq,Sever:2020jjx,Sever:2021nsq,Sever:2021xga}, as well as from arguments based on dual conformal invariance and factorization~\cite{Bern:2008ap}, that the four-particle form factor remainder $R_4$ must reduce to the six-particle MHV amplitude's remainder $\hat{R}_6$ as $T\to0$ in the parametrization introduced in~\eqref{eq:ope_parametrization}.  In both cases, the limit can be interpreted as a triple-collinear limit; in the six-particle case, the limit covers all of the dual-conformally invariant phase space, allowing the triple-collinear splitting amplitude's finite part to be identified with $\hat{R}_6$.  We have checked that this limit is indeed obeyed.

\section{Antipodal Self-Duality}
\label{sec:self-duality}

As discussed in the last section, the four-particle form factor possesses kinematic limits in which it reduces to the three-particle form factor and to the six-particle amplitude. These two quantities were recently discovered to be antipodally dual~\cite{Dixon:2021tdw}, making it tempting to investigate whether the four-particle form factor could be dual to itself in parity-preserving kinematics, where $\text{tr}_5$ vanishes. In the OPE variables, this hypersurface simply corresponds to setting $f_2 = 1$. In the $u_i$ and $v_i$ variables, it requires setting the Gram determinant
\begin{align}
& u_2^2 \bigl[ u_1^2-2 u_1 (1+u_3)+(1-u_3)^2 \bigr]
+ \bigl[ u_1 v_2-v_1 (v_2-u_3) \bigr]^2 \nonumber\\
& - 2 u_2 \bigl[ u_1^2 v_2 + u_1 \bigl(u_3 (2-v_1-v_2)-v_2 (1+v_1) \bigr)
\nonumber\\
&\hskip1cm  + v_1 (1-u_3) (v_2-u_3) \bigr]
\end{align}
to zero.
Note that none of the other four square roots rationalize on this $\text{tr}_5=0$ surface.

\begin{figure}
\includegraphics[width=.8\columnwidth]{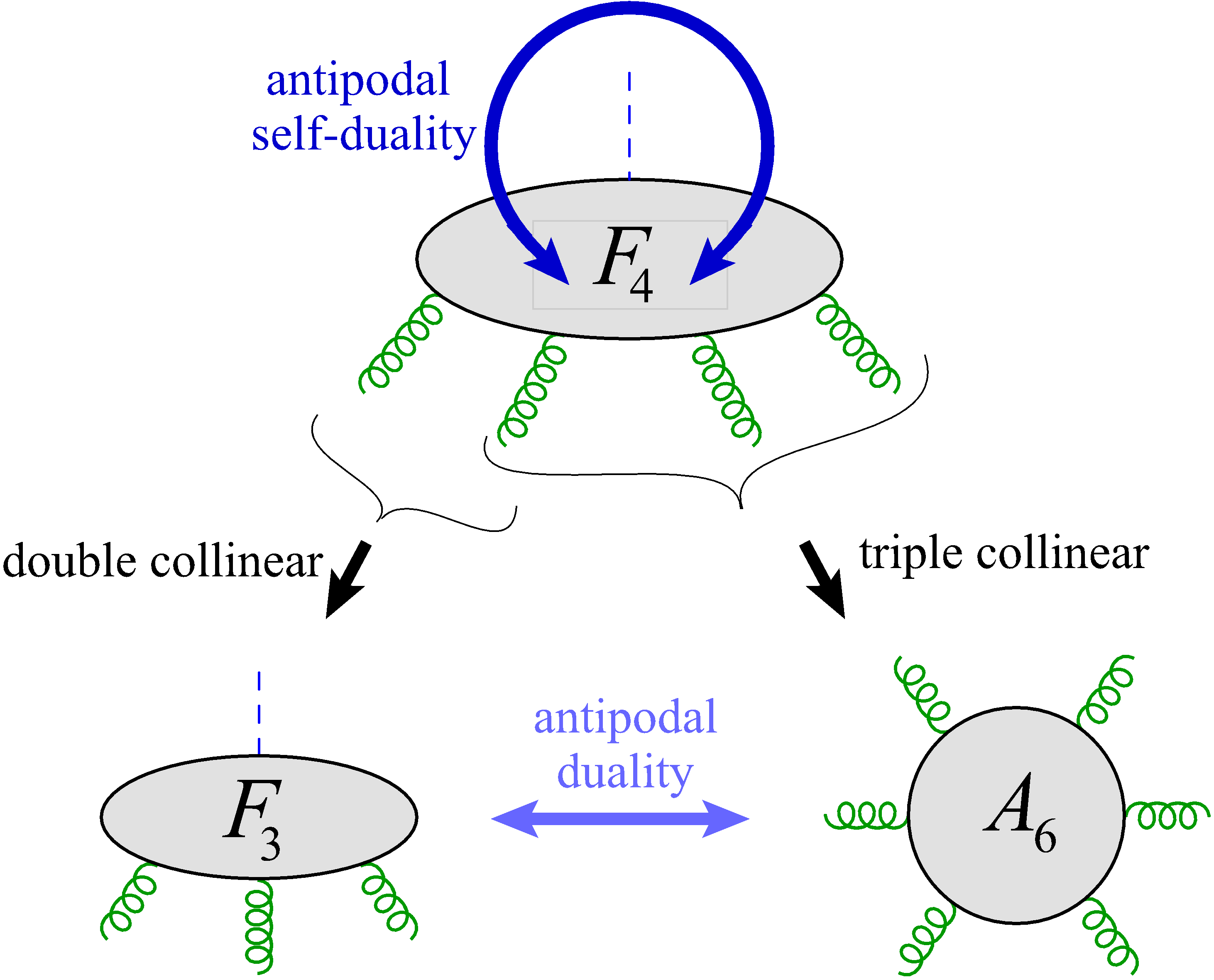}
\caption{The four-particle form factor is antipodally self-dual in parity-preserving kinematics. This duality maps the double and triple collinear limits of the form factor to each other. Because the four-particle form factor reduces to the three-particle form factor and the six-particle amplitude in these limits, the self-duality of the four-particle form factor implies the duality observed in~\cite{Dixon:2021tdw}.}
\label{fig:limits_dualities}
\end{figure}

Surprisingly, we find that an antipodal self-duality does in fact hold on this parity-preserving hypersurface:
 \vspace{-.5cm}
 \begin{equation}
  R_4|_{\text{tr}_5=0}=S\left(R_4|_{\text{tr}_5=0}  
  \right)|_{u_i, \hspace{.01cm} v_i \to g(u_i), \hspace{.03cm} g(v_i)}
\label{eq:self_duality}
 \end{equation}
where $S(F)$ denotes the polylogarithmic antipode of $F$, which acts at the level of the symbol as
 \cite{Gonch3,Brown:2013gia}
 \begin{equation}
 \label{eq: antipode at symbol level}
 S(x_1 \otimes x_2 \otimes \dots \otimes x_m ) = (-1)^m\ x_m \otimes \dots \otimes x_2 \otimes x_1 \, ,
 \end{equation}
and the kinematic map is defined by
\begin{align}
g(u_1) &= u_1 \sqrt{\frac{u_2 u_4}{(u_2-v_1 v_2) (u_4-v_3 v_4)}} \, ,
\label{eq:g_u1} \\
g(v_1) &= (1-v_1) \sqrt{\frac{u_1 u_2}{(u_1-v_4 v_1) (u_2-v_1 v_2)}} \, ,
\label{eq:g_v1}
\end{align}
plus cyclic images. In the OPE variables, this mapping takes an even simpler form:
\begin{equation}
\begin{gathered}
g(T) = \sqrt{\frac{T_2}{S_2}} \, , \quad g(S) = \sqrt{\frac{1}{T_2S_2}} \, , \\
g(T_2) = \frac{T}{S} \, , \quad g(S_2) = \frac{1}{TS} \,, \label{eq:OPE_duality_map}
\end{gathered}
\end{equation}
and it is clear that $g^2=1$.
Notably, this map reduces to the duality map described in~\cite{Dixon:2021tdw} upon identifying $T_2$ and $S_2$ with the OPE parameters that describe the six-particle, which were denoted $\hat{T}$ and $\hat{S}$ in~\cite{Dixon:2021tdw}. This identification naturally arises in the triple-collinear limit of $R_4$, where the form factor reduces to the six-particle remainder $\hat{R}_6$, which is mapped to the double collinear limit of $R_4$ by~\eqref{eq:OPE_duality_map}. However, the self-duality of $R_4$ in~\eqref{eq:self_duality} holds more generally in the {\it full} four-dimensional space of parity-preserving kinematics.\footnote{Note from eqs.~\eqref{eq:g_u1} and \eqref{eq:g_v1} that the light-like limit $u_i,v_i\to\infty$ maps to finite $u_i,v_i$ under the kinematic map $g$, which implies that the light-like form factor does not exhibit the antipodal self-duality we have found for $q^2 \neq 0$.}  Figure~\ref{fig:limits_dualities} depicts the relation between the antipodal self-duality of the four-particle form factor and the previously-observed antipodal duality between form factors and amplitudes.

While antipodal self-duality~\eqref{eq:self_duality} holds for the remainder
function $R_4$, there is an obstruction to it holding for $F_4$.
Namely, the one-loop form factor $\smash{F_4^{(1)}}$ contains final entries that
are not in the set $\{ g(u_i), g(v_i) \}$, $i=1,2,3,4$, dictated by antipodal
self-duality. From the symbol~\eqref{eq:F41} of $\smash{F_4^{(1)}}$, it is easy to
see that the final entry sitting behind the first entry $v_1$ is just
$g(v_1)^2$.  However, the final entry sitting behind $u_1$ is
$(u_1-v_4v_1)/(u_4 u_2)$, and its logarithm is not a linear combination of
$\{ \ln g(u_i), \ln g(v_i) \}$.  Furthermore, it does not seem possible
to repair this obstruction by any simple adjustment of the
normalization $\smash{F_4^{(1)}}$ that is consistent with both the first-entry
condition and the Steinmann relations.

\section{Principle of Maximal Transcendentality}

The two-loop three-particle form factor is known to satisfy the principle of maximal transcendentality (PMT)~\cite{Kotikov:2001sc,Kotikov:2002ab,Kotikov:2004er,Kotikov:2007cy}, meaning that the ${\cal N}=4$ SYM result for $\text{tr}(\phi^2)$ matches the maximally-transcendental part of the Higgs-to-three-gluon amplitude in pure Yang-Mills theory (or QCD) in the leading large-top-mass limit (operator $\text{tr}(F^2)$)~\cite{Wilczek:1977zn,Shifman:1978zn,Dixon:2004za,Gehrmann:2011aa,Brandhuber:2012vm,Duhr:2012fh,Guo:2022pdw}, for all gluon helicity configurations. The PMT has recently been extended (in a different way) to the four-gluon form factor of $\text{tr}(F^3)$~\cite{Guo:2022pdw}. Here we ask: can the PMT for $\text{tr}(\phi^2)$ be extended to any Higgs-to-four-gluon helicity amplitudes?  At one loop, the PMT already fails for the color-ordered helicity configurations $({-}{-}{-}{+})$ and $({-}{+}{-}{+})$, and their parity conjugates~\cite{Glover:2008ffa,Badger:2009hw}, but it works for $({-}{-}{+}{+})$ and $({-}{-}{-}{-})$ and their parity conjugates~\cite{Badger:2007si,Badger:2006us}.

At two loops and leading-color, we cannot say much about $({-}{-}{+}{+})$ currently, because our starting point~\eqref{eq:initial_ansatz} was dihedrally invariant, and the $({-}{-}{+}{+})$ form factor (divided by the tree) need not be invariant, although its leading-transcendental part happens to be at one loop.  That leaves $({-}{-}{-}{-})$, which is dihedrally invariant. The leading transcendental, weight-four, parts of its double collinear limits match those in planar ${\cal N}=4$ SYM, thanks to the PMT holding for the Higgs-to-three-gluon amplitude and for the $g\to gg$ splitting amplitude~\cite{Bern:2004cz}.  We found earlier that there is a unique weight-four, dihedrally- and parity-invariant function with specified double-collinear limits.  Thus $R_4^{(2)}$ should also provide the parity-even part of the two-loop remainder for $A_4^{(2)}(\phi,1^-,2^-,3^-,4^-)$, where $\phi=H+iA$, with $H$ the Higgs boson and $A$ a pseudoscalar coupling to ${\rm tr}(F\tilde{F})$~\cite{Dixon:2004za}.  We also find a unique weight-four parity-odd dihedrally invariant function that vanishes in both the double and triple collinear limits. We leave to future work the tantalizing questions of whether the coefficient of this parity-odd function vanishes, as suggested by the PMT, and more generally, how much of the two-loop Higgs-to-four-gluon amplitude in QCD can be bootstrapped.

\section{Discussion and Conclusions}

In this letter, we have bootstrapped the two-loop four-particle MHV form factor of the chiral part of the stress-tensor supermultiplet in planar $\mathcal{N}=4$ SYM theory, and have found that it possesses an antipodal self-duality in parity-preserving kinematics. While we can only check this duality at two loops, the previously-identified antipodal duality between $R_3$ and $\hat{R}_6$ that it implies has been checked through seven loops (and even eight loops~\cite{ToAppearAmpEightLoop}), which strongly indicates that the self-duality of $R_4$ will hold to all loop orders.

The two-loop four-particle form factor is a highly constrained quantity.  It is determined uniquely by its invariance under Galois and dihedral symmetries, as well as its double-collinear limit. Our result passes a wealth of cross-checks, including the comparison of the near-collinear limit to the FFOPE, triple collinear limits, and the light-like limit. The stringent constraints also raise the question of whether the form factor can be bootstrapped to higher loop orders. The main uncertainty is whether new letters appear at three loops, beyond the 113 letters that appear in the two-loop integrals contributing to this quantity. While great strides have been made recently towards better characterizing the analytic structure of Feynman integrals from first principles (see for instance~\cite{Abreu:2014cla,Bloch:2015efx,Abreu:2017enx,Bourjaily:2020wvq,Hannesdottir:2021kpd,Hannesdottir:2022bmo,Muhlbauer:2022ylo,Hannesdottir:2022xki,Bourjaily:2022vti,Hannesdottir:2022xki,Muhlbauer:2022zzz}), this question remains hard to address without a direct computation.
The answer is that no new letters, beyond the 113, are required to successfully bootstrap the three-loop four-particle MHV form factor; the result is provided in the ancillary file {\bf three\_loop\_symbol.txt}.

It is natural to wonder if further antipodal \mbox{(self-)}dualities hold between form factors and amplitudes at higher particle multiplicity. Notably, as one increases the number of scattering particles, form factors contain an increasingly rich pattern of form factors and amplitudes in their (multi-)collinear limits, giving rise to many intriguing possibilities. In a similar vein, it would be interesting to calculate the two-loop next-to-MHV form form factor for the chiral part of the stress-tensor supermultiplet, and to search for further antipodal dualities that involve this quantity.

More generally, it is critical to understand better the physical reason for the antipodal dualities observed in form factors.  Our work already suggests that the role of the six-point amplitude may be a red herring, that it only participates because it is also the triple-collinear limit of the four-particle form factor.  The simplicity of the map~\eqref{eq:OPE_duality_map} in OPE variables suggests that the reason for the duality might be related to the flux-tube excitations describing the OPE limit. In this context, it is worth noting that there is a fixed surface for~\eqref{eq:OPE_duality_map}, which contains the limit $T, T_2 \to 0$ where small numbers of flux-tube excitations dominate.  This fact may make it possible to examine how the duality acts on single flux-tube excitations.  That action could be a clue in unraveling the physical origin, and thereby predicting in advance where else antipodal duality will emerge.

\vspace{10pt}\acknowledgments

We would like to thank S.~Abreu and V.~Sotnikov for stimulating discussions,
and Gang Yang for helpful comments on the manuscript.
This research was supported by the US Department of Energy under contract
DE--AC02--76SF00515, by the National Science Foundation under Grant
No.~NSF PHY--1748958, and by the Munich Institute for Astro-, Particle
and BioPhysics (MIAPbP) which is funded by the Deutsche
Forschungsgemeinschaft (DFG, German Research Foundation) under Germany's
Excellence Strategy -- EXC--2094 -- 390783311.
LD and YL thank the Kavli Institute for Theoretical Physics,
and LD, AM and MW thank MIAPbP, for support and hospitality.
YL was also supported in part by the Heising-Simons Foundation and the
Simons Foundation. MW was supported in part by the ERC
starting grant 757978 and grant 00025445 from Villum Fonden. \"OG
is supported by the UKRI/EPSRC Stephen Hawking Fellowship EP/T016396/1.
The figure was made with {\sc JaxoDraw}~\cite{Binosi:2008ig}.


\begin{thebibliography}{133}%
\makeatletter
\providecommand \@ifxundefined [1]{%
 \@ifx{#1\undefined}
}%
\providecommand \@ifnum [1]{%
 \ifnum #1\expandafter \@firstoftwo
 \else \expandafter \@secondoftwo
 \fi
}%
\providecommand \@ifx [1]{%
 \ifx #1\expandafter \@firstoftwo
 \else \expandafter \@secondoftwo
 \fi
}%
\providecommand \natexlab [1]{#1}%
\providecommand \enquote  [1]{``#1''}%
\providecommand \bibnamefont  [1]{#1}%
\providecommand \bibfnamefont [1]{#1}%
\providecommand \citenamefont [1]{#1}%
\providecommand \href@noop [0]{\@secondoftwo}%
\providecommand \href [0]{\begingroup \@sanitize@url \@href}%
\providecommand \@href[1]{\@@startlink{#1}\@@href}%
\providecommand \@@href[1]{\endgroup#1\@@endlink}%
\providecommand \@sanitize@url [0]{\catcode `\\12\catcode `\$12\catcode
  `\&12\catcode `\#12\catcode `\^12\catcode `\_12\catcode `\%12\relax}%
\providecommand \@@startlink[1]{}%
\providecommand \@@endlink[0]{}%
\providecommand \url  [0]{\begingroup\@sanitize@url \@url }%
\providecommand \@url [1]{\endgroup\@href {#1}{\urlprefix }}%
\providecommand \urlprefix  [0]{URL }%
\providecommand \Eprint [0]{\href }%
\providecommand \doibase [0]{https://doi.org/}%
\providecommand \selectlanguage [0]{\@gobble}%
\providecommand \bibinfo  [0]{\@secondoftwo}%
\providecommand \bibfield  [0]{\@secondoftwo}%
\providecommand \translation [1]{[#1]}%
\providecommand \BibitemOpen [0]{}%
\providecommand \bibitemStop [0]{}%
\providecommand \bibitemNoStop [0]{.\EOS\space}%
\providecommand \EOS [0]{\spacefactor3000\relax}%
\providecommand \BibitemShut  [1]{\csname bibitem#1\endcsname}%
\let\auto@bib@innerbib\@empty
\bibitem [{\citenamefont {Alday}\ and\ \citenamefont
  {Maldacena}(2007{\natexlab{a}})}]{Alday:2007hr}%
  \BibitemOpen
  \bibfield  {author} {\bibinfo {author} {\bibfnamefont {L.~F.}\ \bibnamefont
  {Alday}}\ and\ \bibinfo {author} {\bibfnamefont {J.~M.}\ \bibnamefont
  {Maldacena}},\ }\bibfield  {title} {\bibinfo {title} {{Gluon scattering
  amplitudes at strong coupling}},\ }\href
  {https://doi.org/10.1088/1126-6708/2007/06/064} {\bibfield  {journal}
  {\bibinfo  {journal} {JHEP}\ }\textbf {\bibinfo {volume} {06}},\ \bibinfo
  {pages} {064}},\ \Eprint {https://arxiv.org/abs/0705.0303} {arXiv:0705.0303
  [hep-th]} \BibitemShut {NoStop}%
\bibitem [{\citenamefont {Drummond}\ \emph {et~al.}(2008)\citenamefont
  {Drummond}, \citenamefont {Korchemsky},\ and\ \citenamefont
  {Sokatchev}}]{Drummond:2007aua}%
  \BibitemOpen
  \bibfield  {author} {\bibinfo {author} {\bibfnamefont {J.}~\bibnamefont
  {Drummond}}, \bibinfo {author} {\bibfnamefont {G.}~\bibnamefont
  {Korchemsky}},\ and\ \bibinfo {author} {\bibfnamefont {E.}~\bibnamefont
  {Sokatchev}},\ }\bibfield  {title} {\bibinfo {title} {{Conformal properties
  of four-gluon planar amplitudes and Wilson loops}},\ }\href
  {https://doi.org/10.1016/j.nuclphysb.2007.11.041} {\bibfield  {journal}
  {\bibinfo  {journal} {Nucl. Phys. B}\ }\textbf {\bibinfo {volume} {795}},\
  \bibinfo {pages} {385} (\bibinfo {year} {2008})},\ \Eprint
  {https://arxiv.org/abs/0707.0243} {arXiv:0707.0243 [hep-th]} \BibitemShut
  {NoStop}%
\bibitem [{\citenamefont {Brandhuber}\ \emph {et~al.}(2008)\citenamefont
  {Brandhuber}, \citenamefont {Heslop},\ and\ \citenamefont
  {Travaglini}}]{Brandhuber:2007yx}%
  \BibitemOpen
  \bibfield  {author} {\bibinfo {author} {\bibfnamefont {A.}~\bibnamefont
  {Brandhuber}}, \bibinfo {author} {\bibfnamefont {P.}~\bibnamefont {Heslop}},\
  and\ \bibinfo {author} {\bibfnamefont {G.}~\bibnamefont {Travaglini}},\
  }\bibfield  {title} {\bibinfo {title} {{MHV amplitudes in $\mathcal{N}=4$
  super Yang-Mills and Wilson loops}},\ }\href
  {https://doi.org/10.1016/j.nuclphysb.2007.11.002} {\bibfield  {journal}
  {\bibinfo  {journal} {Nucl. Phys. B}\ }\textbf {\bibinfo {volume} {794}},\
  \bibinfo {pages} {231} (\bibinfo {year} {2008})},\ \Eprint
  {https://arxiv.org/abs/0707.1153} {arXiv:0707.1153 [hep-th]} \BibitemShut
  {NoStop}%
\bibitem [{\citenamefont {Drummond}\ \emph
  {et~al.}(2010{\natexlab{a}})\citenamefont {Drummond}, \citenamefont {Henn},
  \citenamefont {Korchemsky},\ and\ \citenamefont
  {Sokatchev}}]{Drummond:2007au}%
  \BibitemOpen
  \bibfield  {author} {\bibinfo {author} {\bibfnamefont {J.}~\bibnamefont
  {Drummond}}, \bibinfo {author} {\bibfnamefont {J.}~\bibnamefont {Henn}},
  \bibinfo {author} {\bibfnamefont {G.}~\bibnamefont {Korchemsky}},\ and\
  \bibinfo {author} {\bibfnamefont {E.}~\bibnamefont {Sokatchev}},\ }\bibfield
  {title} {\bibinfo {title} {{Conformal Ward identities for Wilson loops and a
  test of the duality with gluon amplitudes}},\ }\href
  {https://doi.org/10.1016/j.nuclphysb.2009.10.013} {\bibfield  {journal}
  {\bibinfo  {journal} {Nucl.Phys.}\ }\textbf {\bibinfo {volume} {B826}},\
  \bibinfo {pages} {337} (\bibinfo {year} {2010}{\natexlab{a}})},\ \Eprint
  {https://arxiv.org/abs/0712.1223} {arXiv:0712.1223 [hep-th]} \BibitemShut
  {NoStop}%
\bibitem [{\citenamefont {Alday}\ and\ \citenamefont
  {Roiban}(2008)}]{Alday:2008yw}%
  \BibitemOpen
  \bibfield  {author} {\bibinfo {author} {\bibfnamefont {L.~F.}\ \bibnamefont
  {Alday}}\ and\ \bibinfo {author} {\bibfnamefont {R.}~\bibnamefont {Roiban}},\
  }\bibfield  {title} {\bibinfo {title} {{Scattering Amplitudes, Wilson Loops
  and the String/Gauge Theory Correspondence}},\ }\href
  {https://doi.org/10.1016/j.physrep.2008.08.002} {\bibfield  {journal}
  {\bibinfo  {journal} {Phys. Rept.}\ }\textbf {\bibinfo {volume} {468}},\
  \bibinfo {pages} {153} (\bibinfo {year} {2008})},\ \Eprint
  {https://arxiv.org/abs/0807.1889} {arXiv:0807.1889 [hep-th]} \BibitemShut
  {NoStop}%
\bibitem [{\citenamefont {Adamo}\ \emph {et~al.}(2011)\citenamefont {Adamo},
  \citenamefont {Bullimore}, \citenamefont {Mason},\ and\ \citenamefont
  {Skinner}}]{Adamo:2011pv}%
  \BibitemOpen
  \bibfield  {author} {\bibinfo {author} {\bibfnamefont {T.}~\bibnamefont
  {Adamo}}, \bibinfo {author} {\bibfnamefont {M.}~\bibnamefont {Bullimore}},
  \bibinfo {author} {\bibfnamefont {L.}~\bibnamefont {Mason}},\ and\ \bibinfo
  {author} {\bibfnamefont {D.}~\bibnamefont {Skinner}},\ }\bibfield  {title}
  {\bibinfo {title} {{Scattering Amplitudes and Wilson Loops in Twistor
  Space}},\ }\href {https://doi.org/10.1088/1751-8113/44/45/454008} {\bibfield
  {journal} {\bibinfo  {journal} {J. Phys. A}\ }\textbf {\bibinfo {volume}
  {44}},\ \bibinfo {pages} {454008} (\bibinfo {year} {2011})},\ \Eprint
  {https://arxiv.org/abs/1104.2890} {arXiv:1104.2890 [hep-th]} \BibitemShut
  {NoStop}%
\bibitem [{\citenamefont {Brandhuber}\ \emph
  {et~al.}(2011{\natexlab{a}})\citenamefont {Brandhuber}, \citenamefont
  {Spence}, \citenamefont {Travaglini},\ and\ \citenamefont
  {Yang}}]{Brandhuber:2010ad}%
  \BibitemOpen
  \bibfield  {author} {\bibinfo {author} {\bibfnamefont {A.}~\bibnamefont
  {Brandhuber}}, \bibinfo {author} {\bibfnamefont {B.}~\bibnamefont {Spence}},
  \bibinfo {author} {\bibfnamefont {G.}~\bibnamefont {Travaglini}},\ and\
  \bibinfo {author} {\bibfnamefont {G.}~\bibnamefont {Yang}},\ }\bibfield
  {title} {\bibinfo {title} {{Form Factors in $\mathcal{N}=4$ Super Yang-Mills
  and Periodic Wilson Loops}},\ }\href
  {https://doi.org/10.1007/JHEP01(2011)134} {\bibfield  {journal} {\bibinfo
  {journal} {JHEP}\ }\textbf {\bibinfo {volume} {01}},\ \bibinfo {pages}
  {134}},\ \Eprint {https://arxiv.org/abs/1011.1899} {arXiv:1011.1899 [hep-th]}
  \BibitemShut {NoStop}%
\bibitem [{\citenamefont {Alday}\ and\ \citenamefont
  {Maldacena}(2007{\natexlab{b}})}]{Alday:2007he}%
  \BibitemOpen
  \bibfield  {author} {\bibinfo {author} {\bibfnamefont {L.~F.}\ \bibnamefont
  {Alday}}\ and\ \bibinfo {author} {\bibfnamefont {J.}~\bibnamefont
  {Maldacena}},\ }\bibfield  {title} {\bibinfo {title} {{Comments on gluon
  scattering amplitudes via AdS/CFT}},\ }\href
  {https://doi.org/10.1088/1126-6708/2007/11/068} {\bibfield  {journal}
  {\bibinfo  {journal} {JHEP}\ }\textbf {\bibinfo {volume} {0711}},\ \bibinfo
  {pages} {068}},\ \Eprint {https://arxiv.org/abs/0710.1060} {arXiv:0710.1060
  [hep-th]} \BibitemShut {NoStop}%
\bibitem [{\citenamefont {Bern}\ \emph {et~al.}(2008)\citenamefont {Bern},
  \citenamefont {Dixon}, \citenamefont {Kosower}, \citenamefont {Roiban},
  \citenamefont {Spradlin}, \citenamefont {Vergu},\ and\ \citenamefont
  {Volovich}}]{Bern:2008ap}%
  \BibitemOpen
  \bibfield  {author} {\bibinfo {author} {\bibfnamefont {Z.}~\bibnamefont
  {Bern}}, \bibinfo {author} {\bibfnamefont {L.~J.}\ \bibnamefont {Dixon}},
  \bibinfo {author} {\bibfnamefont {D.~A.}\ \bibnamefont {Kosower}}, \bibinfo
  {author} {\bibfnamefont {R.}~\bibnamefont {Roiban}}, \bibinfo {author}
  {\bibfnamefont {M.}~\bibnamefont {Spradlin}}, \bibinfo {author}
  {\bibfnamefont {C.}~\bibnamefont {Vergu}},\ and\ \bibinfo {author}
  {\bibfnamefont {A.}~\bibnamefont {Volovich}},\ }\bibfield  {title} {\bibinfo
  {title} {{The Two-Loop Six-Gluon MHV Amplitude in Maximally Supersymmetric
  Yang-Mills Theory}},\ }\href {https://doi.org/10.1103/PhysRevD.78.045007}
  {\bibfield  {journal} {\bibinfo  {journal} {Phys. Rev.}\ }\textbf {\bibinfo
  {volume} {D78}},\ \bibinfo {pages} {045007} (\bibinfo {year} {2008})},\
  \Eprint {https://arxiv.org/abs/0803.1465} {arXiv:0803.1465 [hep-th]}
  \BibitemShut {NoStop}%
\bibitem [{\citenamefont {Drummond}\ \emph {et~al.}(2009)\citenamefont
  {Drummond}, \citenamefont {Henn}, \citenamefont {Korchemsky},\ and\
  \citenamefont {Sokatchev}}]{Drummond:2008aq}%
  \BibitemOpen
  \bibfield  {author} {\bibinfo {author} {\bibfnamefont {J.}~\bibnamefont
  {Drummond}}, \bibinfo {author} {\bibfnamefont {J.}~\bibnamefont {Henn}},
  \bibinfo {author} {\bibfnamefont {G.}~\bibnamefont {Korchemsky}},\ and\
  \bibinfo {author} {\bibfnamefont {E.}~\bibnamefont {Sokatchev}},\ }\bibfield
  {title} {\bibinfo {title} {{Hexagon Wilson loop = six-gluon MHV amplitude}},\
  }\href {https://doi.org/10.1016/j.nuclphysb.2009.02.015} {\bibfield
  {journal} {\bibinfo  {journal} {Nucl.Phys.}\ }\textbf {\bibinfo {volume}
  {B815}},\ \bibinfo {pages} {142} (\bibinfo {year} {2009})},\ \Eprint
  {https://arxiv.org/abs/0803.1466} {arXiv:0803.1466 [hep-th]} \BibitemShut
  {NoStop}%
\bibitem [{\citenamefont {Maldacena}\ and\ \citenamefont
  {Zhiboedov}(2010)}]{Maldacena:2010kp}%
  \BibitemOpen
  \bibfield  {author} {\bibinfo {author} {\bibfnamefont {J.}~\bibnamefont
  {Maldacena}}\ and\ \bibinfo {author} {\bibfnamefont {A.}~\bibnamefont
  {Zhiboedov}},\ }\bibfield  {title} {\bibinfo {title} {{Form factors at strong
  coupling via a Y-system}},\ }\href {https://doi.org/10.1007/JHEP11(2010)104}
  {\bibfield  {journal} {\bibinfo  {journal} {JHEP}\ }\textbf {\bibinfo
  {volume} {11}},\ \bibinfo {pages} {104}},\ \Eprint
  {https://arxiv.org/abs/1009.1139} {arXiv:1009.1139 [hep-th]} \BibitemShut
  {NoStop}%
\bibitem [{\citenamefont {Ben-Israel}\ \emph {et~al.}(2018)\citenamefont
  {Ben-Israel}, \citenamefont {Tumanov},\ and\ \citenamefont
  {Sever}}]{Ben-Israel:2018ckc}%
  \BibitemOpen
  \bibfield  {author} {\bibinfo {author} {\bibfnamefont {R.}~\bibnamefont
  {Ben-Israel}}, \bibinfo {author} {\bibfnamefont {A.~G.}\ \bibnamefont
  {Tumanov}},\ and\ \bibinfo {author} {\bibfnamefont {A.}~\bibnamefont
  {Sever}},\ }\bibfield  {title} {\bibinfo {title} {{Scattering amplitudes
  \textemdash{} Wilson loops duality for the first non-planar correction}},\
  }\href {https://doi.org/10.1007/JHEP08(2018)122} {\bibfield  {journal}
  {\bibinfo  {journal} {JHEP}\ }\textbf {\bibinfo {volume} {08}},\ \bibinfo
  {pages} {122}},\ \Eprint {https://arxiv.org/abs/1802.09395} {arXiv:1802.09395
  [hep-th]} \BibitemShut {NoStop}%
\bibitem [{\citenamefont {Bianchi}\ \emph {et~al.}(2019)\citenamefont
  {Bianchi}, \citenamefont {Brandhuber}, \citenamefont {Panerai},\ and\
  \citenamefont {Travaglini}}]{Bianchi:2018rrj}%
  \BibitemOpen
  \bibfield  {author} {\bibinfo {author} {\bibfnamefont {L.}~\bibnamefont
  {Bianchi}}, \bibinfo {author} {\bibfnamefont {A.}~\bibnamefont {Brandhuber}},
  \bibinfo {author} {\bibfnamefont {R.}~\bibnamefont {Panerai}},\ and\ \bibinfo
  {author} {\bibfnamefont {G.}~\bibnamefont {Travaglini}},\ }\bibfield  {title}
  {\bibinfo {title} {{Dual conformal invariance for form factors}},\ }\href
  {https://doi.org/10.1007/JHEP02(2019)134} {\bibfield  {journal} {\bibinfo
  {journal} {JHEP}\ }\textbf {\bibinfo {volume} {02}},\ \bibinfo {pages}
  {134}},\ \Eprint {https://arxiv.org/abs/1812.10468} {arXiv:1812.10468
  [hep-th]} \BibitemShut {NoStop}%
\bibitem [{\citenamefont {Drummond}\ \emph {et~al.}(2007)\citenamefont
  {Drummond}, \citenamefont {Henn}, \citenamefont {Smirnov},\ and\
  \citenamefont {Sokatchev}}]{Drummond:2006rz}%
  \BibitemOpen
  \bibfield  {author} {\bibinfo {author} {\bibfnamefont {J.~M.}\ \bibnamefont
  {Drummond}}, \bibinfo {author} {\bibfnamefont {J.}~\bibnamefont {Henn}},
  \bibinfo {author} {\bibfnamefont {V.~A.}\ \bibnamefont {Smirnov}},\ and\
  \bibinfo {author} {\bibfnamefont {E.}~\bibnamefont {Sokatchev}},\ }\bibfield
  {title} {\bibinfo {title} {{Magic identities for conformal four-point
  integrals}},\ }\href {https://doi.org/10.1088/1126-6708/2007/01/064}
  {\bibfield  {journal} {\bibinfo  {journal} {JHEP}\ }\textbf {\bibinfo
  {volume} {01}},\ \bibinfo {pages} {064}},\ \Eprint
  {https://arxiv.org/abs/hep-th/0607160} {arXiv:hep-th/0607160 [hep-th]}
  \BibitemShut {NoStop}%
\bibitem [{\citenamefont {Bern}\ \emph
  {et~al.}(2007{\natexlab{a}})\citenamefont {Bern}, \citenamefont {Czakon},
  \citenamefont {Dixon}, \citenamefont {Kosower},\ and\ \citenamefont
  {Smirnov}}]{Bern:2006ew}%
  \BibitemOpen
  \bibfield  {author} {\bibinfo {author} {\bibfnamefont {Z.}~\bibnamefont
  {Bern}}, \bibinfo {author} {\bibfnamefont {M.}~\bibnamefont {Czakon}},
  \bibinfo {author} {\bibfnamefont {L.~J.}\ \bibnamefont {Dixon}}, \bibinfo
  {author} {\bibfnamefont {D.~A.}\ \bibnamefont {Kosower}},\ and\ \bibinfo
  {author} {\bibfnamefont {V.~A.}\ \bibnamefont {Smirnov}},\ }\bibfield
  {title} {\bibinfo {title} {{The Four-Loop Planar Amplitude and Cusp Anomalous
  Dimension in Maximally Supersymmetric Yang-Mills Theory}},\ }\href
  {https://doi.org/10.1103/PhysRevD.75.085010} {\bibfield  {journal} {\bibinfo
  {journal} {Phys. Rev.}\ }\textbf {\bibinfo {volume} {D75}},\ \bibinfo {pages}
  {085010} (\bibinfo {year} {2007}{\natexlab{a}})},\ \Eprint
  {https://arxiv.org/abs/hep-th/0610248} {arXiv:hep-th/0610248 [hep-th]}
  \BibitemShut {NoStop}%
\bibitem [{\citenamefont {Bern}\ \emph
  {et~al.}(2007{\natexlab{b}})\citenamefont {Bern}, \citenamefont {Carrasco},
  \citenamefont {Johansson},\ and\ \citenamefont {Kosower}}]{Bern:2007ct}%
  \BibitemOpen
  \bibfield  {author} {\bibinfo {author} {\bibfnamefont {Z.}~\bibnamefont
  {Bern}}, \bibinfo {author} {\bibfnamefont {J.}~\bibnamefont {Carrasco}},
  \bibinfo {author} {\bibfnamefont {H.}~\bibnamefont {Johansson}},\ and\
  \bibinfo {author} {\bibfnamefont {D.}~\bibnamefont {Kosower}},\ }\bibfield
  {title} {\bibinfo {title} {{Maximally supersymmetric planar Yang-Mills
  amplitudes at five loops}},\ }\href
  {https://doi.org/10.1103/PhysRevD.76.125020} {\bibfield  {journal} {\bibinfo
  {journal} {Phys.Rev.}\ }\textbf {\bibinfo {volume} {D76}},\ \bibinfo {pages}
  {125020} (\bibinfo {year} {2007}{\natexlab{b}})},\ \Eprint
  {https://arxiv.org/abs/0705.1864} {arXiv:0705.1864 [hep-th]} \BibitemShut
  {NoStop}%
\bibitem [{\citenamefont {Drummond}\ \emph
  {et~al.}(2010{\natexlab{b}})\citenamefont {Drummond}, \citenamefont {Henn},
  \citenamefont {Korchemsky},\ and\ \citenamefont
  {Sokatchev}}]{Drummond:2008vq}%
  \BibitemOpen
  \bibfield  {author} {\bibinfo {author} {\bibfnamefont {J.~M.}\ \bibnamefont
  {Drummond}}, \bibinfo {author} {\bibfnamefont {J.}~\bibnamefont {Henn}},
  \bibinfo {author} {\bibfnamefont {G.~P.}\ \bibnamefont {Korchemsky}},\ and\
  \bibinfo {author} {\bibfnamefont {E.}~\bibnamefont {Sokatchev}},\ }\bibfield
  {title} {\bibinfo {title} {{Dual superconformal symmetry of scattering
  amplitudes in $\mathcal{N}=4$ super-Yang-Mills theory}},\ }\href
  {https://doi.org/10.1016/j.nuclphysb.2009.11.022} {\bibfield  {journal}
  {\bibinfo  {journal} {Nucl. Phys.}\ }\textbf {\bibinfo {volume} {B828}},\
  \bibinfo {pages} {317} (\bibinfo {year} {2010}{\natexlab{b}})},\ \Eprint
  {https://arxiv.org/abs/0807.1095} {arXiv:0807.1095 [hep-th]} \BibitemShut
  {NoStop}%
\bibitem [{\citenamefont {Caron-Huot}\ \emph
  {et~al.}(2019{\natexlab{a}})\citenamefont {Caron-Huot}, \citenamefont
  {Dixon}, \citenamefont {Dulat}, \citenamefont {Von~Hippel}, \citenamefont
  {McLeod},\ and\ \citenamefont {Papathanasiou}}]{Caron-Huot:2019bsq}%
  \BibitemOpen
  \bibfield  {author} {\bibinfo {author} {\bibfnamefont {S.}~\bibnamefont
  {Caron-Huot}}, \bibinfo {author} {\bibfnamefont {L.~J.}\ \bibnamefont
  {Dixon}}, \bibinfo {author} {\bibfnamefont {F.}~\bibnamefont {Dulat}},
  \bibinfo {author} {\bibfnamefont {M.}~\bibnamefont {Von~Hippel}}, \bibinfo
  {author} {\bibfnamefont {A.~J.}\ \bibnamefont {McLeod}},\ and\ \bibinfo
  {author} {\bibfnamefont {G.}~\bibnamefont {Papathanasiou}},\ }\bibfield
  {title} {\bibinfo {title} {{The Cosmic Galois Group and Extended Steinmann
  Relations for Planar $\mathcal{N} = 4$ SYM Amplitudes}},\ }\href
  {https://doi.org/10.1007/JHEP09(2019)061} {\bibfield  {journal} {\bibinfo
  {journal} {JHEP}\ }\textbf {\bibinfo {volume} {09}},\ \bibinfo {pages}
  {061}},\ \Eprint {https://arxiv.org/abs/1906.07116} {arXiv:1906.07116
  [hep-th]} \BibitemShut {NoStop}%
\bibitem [{\citenamefont {Golden}\ \emph
  {et~al.}(2014{\natexlab{a}})\citenamefont {Golden}, \citenamefont
  {Goncharov}, \citenamefont {Spradlin}, \citenamefont {Vergu},\ and\
  \citenamefont {Volovich}}]{Golden:2013xva}%
  \BibitemOpen
  \bibfield  {author} {\bibinfo {author} {\bibfnamefont {J.}~\bibnamefont
  {Golden}}, \bibinfo {author} {\bibfnamefont {A.~B.}\ \bibnamefont
  {Goncharov}}, \bibinfo {author} {\bibfnamefont {M.}~\bibnamefont {Spradlin}},
  \bibinfo {author} {\bibfnamefont {C.}~\bibnamefont {Vergu}},\ and\ \bibinfo
  {author} {\bibfnamefont {A.}~\bibnamefont {Volovich}},\ }\bibfield  {title}
  {\bibinfo {title} {{Motivic Amplitudes and Cluster Coordinates}},\ }\href
  {https://doi.org/10.1007/JHEP01(2014)091} {\bibfield  {journal} {\bibinfo
  {journal} {JHEP}\ }\textbf {\bibinfo {volume} {1401}},\ \bibinfo {pages}
  {091}},\ \Eprint {https://arxiv.org/abs/1305.1617} {arXiv:1305.1617 [hep-th]}
  \BibitemShut {NoStop}%
\bibitem [{\citenamefont {Golden}\ and\ \citenamefont
  {Spradlin}(2015)}]{Golden:2014pua}%
  \BibitemOpen
  \bibfield  {author} {\bibinfo {author} {\bibfnamefont {J.}~\bibnamefont
  {Golden}}\ and\ \bibinfo {author} {\bibfnamefont {M.}~\bibnamefont
  {Spradlin}},\ }\bibfield  {title} {\bibinfo {title} {{A Cluster Bootstrap for
  Two-Loop MHV Amplitudes}},\ }\href {https://doi.org/10.1007/JHEP02(2015)002}
  {\bibfield  {journal} {\bibinfo  {journal} {JHEP}\ }\textbf {\bibinfo
  {volume} {02}},\ \bibinfo {pages} {002}},\ \Eprint
  {https://arxiv.org/abs/1411.3289} {arXiv:1411.3289 [hep-th]} \BibitemShut
  {NoStop}%
\bibitem [{\citenamefont {Golden}\ \emph
  {et~al.}(2014{\natexlab{b}})\citenamefont {Golden}, \citenamefont {Paulos},
  \citenamefont {Spradlin},\ and\ \citenamefont {Volovich}}]{Golden:2014xqa}%
  \BibitemOpen
  \bibfield  {author} {\bibinfo {author} {\bibfnamefont {J.}~\bibnamefont
  {Golden}}, \bibinfo {author} {\bibfnamefont {M.~F.}\ \bibnamefont {Paulos}},
  \bibinfo {author} {\bibfnamefont {M.}~\bibnamefont {Spradlin}},\ and\
  \bibinfo {author} {\bibfnamefont {A.}~\bibnamefont {Volovich}},\ }\bibfield
  {title} {\bibinfo {title} {{Cluster Polylogarithms for Scattering
  Amplitudes}},\ }\href {https://doi.org/10.1088/1751-8113/47/47/474005}
  {\bibfield  {journal} {\bibinfo  {journal} {J. Phys. A}\ }\textbf {\bibinfo
  {volume} {47}},\ \bibinfo {pages} {474005} (\bibinfo {year}
  {2014}{\natexlab{b}})},\ \Eprint {https://arxiv.org/abs/1401.6446}
  {arXiv:1401.6446 [hep-th]} \BibitemShut {NoStop}%
\bibitem [{\citenamefont {Golden}\ and\ \citenamefont
  {Spradlin}(2014)}]{Golden:2014xqf}%
  \BibitemOpen
  \bibfield  {author} {\bibinfo {author} {\bibfnamefont {J.}~\bibnamefont
  {Golden}}\ and\ \bibinfo {author} {\bibfnamefont {M.}~\bibnamefont
  {Spradlin}},\ }\bibfield  {title} {\bibinfo {title} {{An analytic result for
  the two-loop seven-point MHV amplitude in $ \mathcal{N} $ = 4 SYM}},\ }\href
  {https://doi.org/10.1007/JHEP08(2014)154} {\bibfield  {journal} {\bibinfo
  {journal} {JHEP}\ }\textbf {\bibinfo {volume} {1408}},\ \bibinfo {pages}
  {154}},\ \Eprint {https://arxiv.org/abs/1406.2055} {arXiv:1406.2055 [hep-th]}
  \BibitemShut {NoStop}%
\bibitem [{\citenamefont {Drummond}\ \emph {et~al.}(2018)\citenamefont
  {Drummond}, \citenamefont {Foster},\ and\ \citenamefont
  {G{\"u}rdo{\u{g}}an}}]{Drummond:2017ssj}%
  \BibitemOpen
  \bibfield  {author} {\bibinfo {author} {\bibfnamefont {J.}~\bibnamefont
  {Drummond}}, \bibinfo {author} {\bibfnamefont {J.}~\bibnamefont {Foster}},\
  and\ \bibinfo {author} {\bibfnamefont {{\"O}.}~\bibnamefont
  {G{\"u}rdo{\u{g}}an}},\ }\bibfield  {title} {\bibinfo {title} {{Cluster
  Adjacency Properties of Scattering Amplitudes in $\mathcal{N}=4$
  Supersymmetric Yang-Mills Theory}},\ }\href
  {https://doi.org/10.1103/PhysRevLett.120.161601} {\bibfield  {journal}
  {\bibinfo  {journal} {Phys. Rev. Lett.}\ }\textbf {\bibinfo {volume} {120}},\
  \bibinfo {pages} {161601} (\bibinfo {year} {2018})},\ \Eprint
  {https://arxiv.org/abs/1710.10953} {arXiv:1710.10953 [hep-th]} \BibitemShut
  {NoStop}%
\bibitem [{\citenamefont {Bourjaily}\ \emph {et~al.}(2018)\citenamefont
  {Bourjaily}, \citenamefont {McLeod}, \citenamefont {von Hippel},\ and\
  \citenamefont {Wilhelm}}]{Bourjaily:2018aeq}%
  \BibitemOpen
  \bibfield  {author} {\bibinfo {author} {\bibfnamefont {J.~L.}\ \bibnamefont
  {Bourjaily}}, \bibinfo {author} {\bibfnamefont {A.~J.}\ \bibnamefont
  {McLeod}}, \bibinfo {author} {\bibfnamefont {M.}~\bibnamefont {von Hippel}},\
  and\ \bibinfo {author} {\bibfnamefont {M.}~\bibnamefont {Wilhelm}},\
  }\bibfield  {title} {\bibinfo {title} {{Rationalizing Loop Integration}},\
  }\href {https://doi.org/10.1007/JHEP08(2018)184} {\bibfield  {journal}
  {\bibinfo  {journal} {JHEP}\ }\textbf {\bibinfo {volume} {08}},\ \bibinfo
  {pages} {184}},\ \Eprint {https://arxiv.org/abs/1805.10281} {arXiv:1805.10281
  [hep-th]} \BibitemShut {NoStop}%
\bibitem [{\citenamefont {Drummond}\ \emph
  {et~al.}(2019{\natexlab{a}})\citenamefont {Drummond}, \citenamefont
  {Foster},\ and\ \citenamefont {G{\"u}rdo{\u{g}}an}}]{Drummond:2018dfd}%
  \BibitemOpen
  \bibfield  {author} {\bibinfo {author} {\bibfnamefont {J.}~\bibnamefont
  {Drummond}}, \bibinfo {author} {\bibfnamefont {J.}~\bibnamefont {Foster}},\
  and\ \bibinfo {author} {\bibfnamefont {{\"O}.}~\bibnamefont
  {G{\"u}rdo{\u{g}}an}},\ }\bibfield  {title} {\bibinfo {title} {{Cluster
  adjacency beyond MHV}},\ }\href {https://doi.org/10.1007/JHEP03(2019)086}
  {\bibfield  {journal} {\bibinfo  {journal} {JHEP}\ }\textbf {\bibinfo
  {volume} {03}},\ \bibinfo {pages} {086}},\ \Eprint
  {https://arxiv.org/abs/1810.08149} {arXiv:1810.08149 [hep-th]} \BibitemShut
  {NoStop}%
\bibitem [{\citenamefont {Golden}\ and\ \citenamefont
  {McLeod}(2019)}]{Golden:2018gtk}%
  \BibitemOpen
  \bibfield  {author} {\bibinfo {author} {\bibfnamefont {J.}~\bibnamefont
  {Golden}}\ and\ \bibinfo {author} {\bibfnamefont {A.~J.}\ \bibnamefont
  {McLeod}},\ }\bibfield  {title} {\bibinfo {title} {{Cluster Algebras and the
  Subalgebra Constructibility of the Seven-Particle Remainder Function}},\
  }\href {https://doi.org/10.1007/JHEP01(2019)017} {\bibfield  {journal}
  {\bibinfo  {journal} {JHEP}\ }\textbf {\bibinfo {volume} {01}},\ \bibinfo
  {pages} {017}},\ \Eprint {https://arxiv.org/abs/1810.12181} {arXiv:1810.12181
  [hep-th]} \BibitemShut {NoStop}%
\bibitem [{\citenamefont {Drummond}\ \emph
  {et~al.}(2019{\natexlab{b}})\citenamefont {Drummond}, \citenamefont {Foster},
  \citenamefont {G{\"u}rdo{\u{g}}an},\ and\ \citenamefont
  {Papathanasiou}}]{Drummond:2018caf}%
  \BibitemOpen
  \bibfield  {author} {\bibinfo {author} {\bibfnamefont {J.}~\bibnamefont
  {Drummond}}, \bibinfo {author} {\bibfnamefont {J.}~\bibnamefont {Foster}},
  \bibinfo {author} {\bibfnamefont {{\"O}.}~\bibnamefont
  {G{\"u}rdo{\u{g}}an}},\ and\ \bibinfo {author} {\bibfnamefont
  {G.}~\bibnamefont {Papathanasiou}},\ }\bibfield  {title} {\bibinfo {title}
  {{Cluster adjacency and the four-loop NMHV heptagon}},\ }\href
  {https://doi.org/10.1007/JHEP03(2019)087} {\bibfield  {journal} {\bibinfo
  {journal} {JHEP}\ }\textbf {\bibinfo {volume} {03}},\ \bibinfo {pages}
  {087}},\ \Eprint {https://arxiv.org/abs/1812.04640} {arXiv:1812.04640
  [hep-th]} \BibitemShut {NoStop}%
\bibitem [{\citenamefont {Golden}\ \emph {et~al.}(2019)\citenamefont {Golden},
  \citenamefont {McLeod}, \citenamefont {Spradlin},\ and\ \citenamefont
  {Volovich}}]{Golden:2019kks}%
  \BibitemOpen
  \bibfield  {author} {\bibinfo {author} {\bibfnamefont {J.}~\bibnamefont
  {Golden}}, \bibinfo {author} {\bibfnamefont {A.~J.}\ \bibnamefont {McLeod}},
  \bibinfo {author} {\bibfnamefont {M.}~\bibnamefont {Spradlin}},\ and\
  \bibinfo {author} {\bibfnamefont {A.}~\bibnamefont {Volovich}},\ }\bibfield
  {title} {\bibinfo {title} {{The Sklyanin Bracket and Cluster Adjacency at All
  Multiplicity}},\ }\href {https://doi.org/10.1007/JHEP03(2019)195} {\bibfield
  {journal} {\bibinfo  {journal} {JHEP}\ }\textbf {\bibinfo {volume} {03}},\
  \bibinfo {pages} {195}},\ \Eprint {https://arxiv.org/abs/1902.11286}
  {arXiv:1902.11286 [hep-th]} \BibitemShut {NoStop}%
\bibitem [{\citenamefont {Golden}\ and\ \citenamefont
  {McLeod}(2021)}]{Golden:2021ggj}%
  \BibitemOpen
  \bibfield  {author} {\bibinfo {author} {\bibfnamefont {J.}~\bibnamefont
  {Golden}}\ and\ \bibinfo {author} {\bibfnamefont {A.~J.}\ \bibnamefont
  {McLeod}},\ }\bibfield  {title} {\bibinfo {title} {{The two-loop remainder
  function for eight and nine particles}},\ }\href
  {https://doi.org/10.1007/JHEP06(2021)142} {\bibfield  {journal} {\bibinfo
  {journal} {JHEP}\ }\textbf {\bibinfo {volume} {06}},\ \bibinfo {pages}
  {142}},\ \Eprint {https://arxiv.org/abs/2104.14194} {arXiv:2104.14194
  [hep-th]} \BibitemShut {NoStop}%
\bibitem [{\citenamefont {Drummond}\ \emph {et~al.}(2020)\citenamefont
  {Drummond}, \citenamefont {Foster}, \citenamefont {G{\"u}rdo{\u{g}}an},\ and\
  \citenamefont {Kalousios}}]{Drummond:2019qjk}%
  \BibitemOpen
  \bibfield  {author} {\bibinfo {author} {\bibfnamefont {J.}~\bibnamefont
  {Drummond}}, \bibinfo {author} {\bibfnamefont {J.}~\bibnamefont {Foster}},
  \bibinfo {author} {\bibfnamefont {{\"O}.}~\bibnamefont
  {G{\"u}rdo{\u{g}}an}},\ and\ \bibinfo {author} {\bibfnamefont
  {C.}~\bibnamefont {Kalousios}},\ }\bibfield  {title} {\bibinfo {title}
  {{Tropical Grassmannians, cluster algebras and scattering amplitudes}},\
  }\href {https://doi.org/10.1007/JHEP04(2020)146} {\bibfield  {journal}
  {\bibinfo  {journal} {JHEP}\ }\textbf {\bibinfo {volume} {04}},\ \bibinfo
  {pages} {146}},\ \Eprint {https://arxiv.org/abs/1907.01053} {arXiv:1907.01053
  [hep-th]} \BibitemShut {NoStop}%
\bibitem [{\citenamefont {Drummond}\ \emph
  {et~al.}(2021{\natexlab{a}})\citenamefont {Drummond}, \citenamefont {Foster},
  \citenamefont {G{\"u}rdo{\u{g}}an},\ and\ \citenamefont
  {Kalousios}}]{Drummond:2019cxm}%
  \BibitemOpen
  \bibfield  {author} {\bibinfo {author} {\bibfnamefont {J.}~\bibnamefont
  {Drummond}}, \bibinfo {author} {\bibfnamefont {J.}~\bibnamefont {Foster}},
  \bibinfo {author} {\bibfnamefont {{\"O}.}~\bibnamefont
  {G{\"u}rdo{\u{g}}an}},\ and\ \bibinfo {author} {\bibfnamefont
  {C.}~\bibnamefont {Kalousios}},\ }\bibfield  {title} {\bibinfo {title}
  {{Algebraic singularities of scattering amplitudes from tropical geometry}},\
  }\href {https://doi.org/10.1007/JHEP04(2021)002} {\bibfield  {journal}
  {\bibinfo  {journal} {JHEP}\ }\textbf {\bibinfo {volume} {04}},\ \bibinfo
  {pages} {002}},\ \Eprint {https://arxiv.org/abs/1912.08217} {arXiv:1912.08217
  [hep-th]} \BibitemShut {NoStop}%
\bibitem [{\citenamefont {Arkani-Hamed}\ \emph {et~al.}(2021)\citenamefont
  {Arkani-Hamed}, \citenamefont {Lam},\ and\ \citenamefont
  {Spradlin}}]{Arkani-Hamed:2019rds}%
  \BibitemOpen
  \bibfield  {author} {\bibinfo {author} {\bibfnamefont {N.}~\bibnamefont
  {Arkani-Hamed}}, \bibinfo {author} {\bibfnamefont {T.}~\bibnamefont {Lam}},\
  and\ \bibinfo {author} {\bibfnamefont {M.}~\bibnamefont {Spradlin}},\
  }\bibfield  {title} {\bibinfo {title} {{Non-perturbative geometries for
  planar $\mathcal{N}=4$ SYM amplitudes}},\ }\href
  {https://doi.org/10.1007/JHEP03(2021)065} {\bibfield  {journal} {\bibinfo
  {journal} {JHEP}\ }\textbf {\bibinfo {volume} {03}},\ \bibinfo {pages}
  {065}},\ \Eprint {https://arxiv.org/abs/1912.08222} {arXiv:1912.08222
  [hep-th]} \BibitemShut {NoStop}%
\bibitem [{\citenamefont {Henke}\ and\ \citenamefont
  {Papathanasiou}(2020)}]{Henke:2019hve}%
  \BibitemOpen
  \bibfield  {author} {\bibinfo {author} {\bibfnamefont {N.}~\bibnamefont
  {Henke}}\ and\ \bibinfo {author} {\bibfnamefont {G.}~\bibnamefont
  {Papathanasiou}},\ }\bibfield  {title} {\bibinfo {title} {{How tropical are
  seven- and eight-particle amplitudes?}},\ }\href
  {https://doi.org/10.1007/JHEP08(2020)005} {\bibfield  {journal} {\bibinfo
  {journal} {JHEP}\ }\textbf {\bibinfo {volume} {08}},\ \bibinfo {pages}
  {005}},\ \Eprint {https://arxiv.org/abs/1912.08254} {arXiv:1912.08254
  [hep-th]} \BibitemShut {NoStop}%
\bibitem [{\citenamefont {Drummond}\ \emph
  {et~al.}(2021{\natexlab{b}})\citenamefont {Drummond}, \citenamefont {Foster},
  \citenamefont {G{\"u}rdo{\u{g}}an},\ and\ \citenamefont
  {Kalousios}}]{Drummond:2020kqg}%
  \BibitemOpen
  \bibfield  {author} {\bibinfo {author} {\bibfnamefont {J.}~\bibnamefont
  {Drummond}}, \bibinfo {author} {\bibfnamefont {J.}~\bibnamefont {Foster}},
  \bibinfo {author} {\bibfnamefont {{\"O}.}~\bibnamefont
  {G{\"u}rdo{\u{g}}an}},\ and\ \bibinfo {author} {\bibfnamefont
  {C.}~\bibnamefont {Kalousios}},\ }\bibfield  {title} {\bibinfo {title}
  {{Tropical fans, scattering equations and amplitudes}},\ }\href
  {https://doi.org/10.1007/JHEP11(2021)071} {\bibfield  {journal} {\bibinfo
  {journal} {JHEP}\ }\textbf {\bibinfo {volume} {11}},\ \bibinfo {pages}
  {071}},\ \Eprint {https://arxiv.org/abs/2002.04624} {arXiv:2002.04624
  [hep-th]} \BibitemShut {NoStop}%
\bibitem [{\citenamefont {Mago}\ \emph {et~al.}(2020)\citenamefont {Mago},
  \citenamefont {Schreiber}, \citenamefont {Spradlin},\ and\ \citenamefont
  {Volovich}}]{Mago:2020kmp}%
  \BibitemOpen
  \bibfield  {author} {\bibinfo {author} {\bibfnamefont {J.}~\bibnamefont
  {Mago}}, \bibinfo {author} {\bibfnamefont {A.}~\bibnamefont {Schreiber}},
  \bibinfo {author} {\bibfnamefont {M.}~\bibnamefont {Spradlin}},\ and\
  \bibinfo {author} {\bibfnamefont {A.}~\bibnamefont {Volovich}},\ }\bibfield
  {title} {\bibinfo {title} {{Symbol alphabets from plabic graphs}},\ }\href
  {https://doi.org/10.1007/JHEP10(2020)128} {\bibfield  {journal} {\bibinfo
  {journal} {JHEP}\ }\textbf {\bibinfo {volume} {10}},\ \bibinfo {pages}
  {128}},\ \Eprint {https://arxiv.org/abs/2007.00646} {arXiv:2007.00646
  [hep-th]} \BibitemShut {NoStop}%
\bibitem [{\citenamefont {Chicherin}\ \emph {et~al.}(2021)\citenamefont
  {Chicherin}, \citenamefont {Henn},\ and\ \citenamefont
  {Papathanasiou}}]{Chicherin:2020umh}%
  \BibitemOpen
  \bibfield  {author} {\bibinfo {author} {\bibfnamefont {D.}~\bibnamefont
  {Chicherin}}, \bibinfo {author} {\bibfnamefont {J.~M.}\ \bibnamefont
  {Henn}},\ and\ \bibinfo {author} {\bibfnamefont {G.}~\bibnamefont
  {Papathanasiou}},\ }\bibfield  {title} {\bibinfo {title} {{Cluster algebras
  for Feynman integrals}},\ }\href
  {https://doi.org/10.1103/PhysRevLett.126.091603} {\bibfield  {journal}
  {\bibinfo  {journal} {Phys. Rev. Lett.}\ }\textbf {\bibinfo {volume} {126}},\
  \bibinfo {pages} {091603} (\bibinfo {year} {2021})},\ \Eprint
  {https://arxiv.org/abs/2012.12285} {arXiv:2012.12285 [hep-th]} \BibitemShut
  {NoStop}%
\bibitem [{\citenamefont {Mago}\ \emph
  {et~al.}(2021{\natexlab{a}})\citenamefont {Mago}, \citenamefont {Schreiber},
  \citenamefont {Spradlin}, \citenamefont {Yelleshpur~Srikant},\ and\
  \citenamefont {Volovich}}]{Mago:2020nuv}%
  \BibitemOpen
  \bibfield  {author} {\bibinfo {author} {\bibfnamefont {J.}~\bibnamefont
  {Mago}}, \bibinfo {author} {\bibfnamefont {A.}~\bibnamefont {Schreiber}},
  \bibinfo {author} {\bibfnamefont {M.}~\bibnamefont {Spradlin}}, \bibinfo
  {author} {\bibfnamefont {A.}~\bibnamefont {Yelleshpur~Srikant}},\ and\
  \bibinfo {author} {\bibfnamefont {A.}~\bibnamefont {Volovich}},\ }\bibfield
  {title} {\bibinfo {title} {{Symbol alphabets from plabic graphs II: rational
  letters}},\ }\href {https://doi.org/10.1007/JHEP04(2021)056} {\bibfield
  {journal} {\bibinfo  {journal} {JHEP}\ }\textbf {\bibinfo {volume} {04}},\
  \bibinfo {pages} {056}},\ \Eprint {https://arxiv.org/abs/2012.15812}
  {arXiv:2012.15812 [hep-th]} \BibitemShut {NoStop}%
\bibitem [{\citenamefont {Herderschee}(2021)}]{Herderschee:2021dez}%
  \BibitemOpen
  \bibfield  {author} {\bibinfo {author} {\bibfnamefont {A.}~\bibnamefont
  {Herderschee}},\ }\bibfield  {title} {\bibinfo {title} {{Algebraic branch
  points at all loop orders from positive kinematics and wall crossing}},\
  }\href {https://doi.org/10.1007/JHEP07(2021)049} {\bibfield  {journal}
  {\bibinfo  {journal} {JHEP}\ }\textbf {\bibinfo {volume} {07}},\ \bibinfo
  {pages} {049}},\ \Eprint {https://arxiv.org/abs/2102.03611} {arXiv:2102.03611
  [hep-th]} \BibitemShut {NoStop}%
\bibitem [{\citenamefont {He}\ \emph {et~al.}(2021{\natexlab{a}})\citenamefont
  {He}, \citenamefont {Li},\ and\ \citenamefont {Yang}}]{He:2021esx}%
  \BibitemOpen
  \bibfield  {author} {\bibinfo {author} {\bibfnamefont {S.}~\bibnamefont
  {He}}, \bibinfo {author} {\bibfnamefont {Z.}~\bibnamefont {Li}},\ and\
  \bibinfo {author} {\bibfnamefont {Q.}~\bibnamefont {Yang}},\ }\bibfield
  {title} {\bibinfo {title} {{Notes on cluster algebras and some all-loop
  Feynman integrals}},\ }\href {https://doi.org/10.1007/JHEP06(2021)119}
  {\bibfield  {journal} {\bibinfo  {journal} {JHEP}\ }\textbf {\bibinfo
  {volume} {06}},\ \bibinfo {pages} {119}},\ \Eprint
  {https://arxiv.org/abs/2103.02796} {arXiv:2103.02796 [hep-th]} \BibitemShut
  {NoStop}%
\bibitem [{\citenamefont {Mago}\ \emph
  {et~al.}(2021{\natexlab{b}})\citenamefont {Mago}, \citenamefont {Schreiber},
  \citenamefont {Spradlin}, \citenamefont {Yelleshpur~Srikant},\ and\
  \citenamefont {Volovich}}]{Mago:2021luw}%
  \BibitemOpen
  \bibfield  {author} {\bibinfo {author} {\bibfnamefont {J.}~\bibnamefont
  {Mago}}, \bibinfo {author} {\bibfnamefont {A.}~\bibnamefont {Schreiber}},
  \bibinfo {author} {\bibfnamefont {M.}~\bibnamefont {Spradlin}}, \bibinfo
  {author} {\bibfnamefont {A.}~\bibnamefont {Yelleshpur~Srikant}},\ and\
  \bibinfo {author} {\bibfnamefont {A.}~\bibnamefont {Volovich}},\ }\bibfield
  {title} {\bibinfo {title} {{Symbol alphabets from plabic graphs III: $n =
  9$}},\ }\href {https://doi.org/10.1007/JHEP09(2021)002} {\bibfield  {journal}
  {\bibinfo  {journal} {JHEP}\ }\textbf {\bibinfo {volume} {09}},\ \bibinfo
  {pages} {002}},\ \Eprint {https://arxiv.org/abs/2106.01406} {arXiv:2106.01406
  [hep-th]} \BibitemShut {NoStop}%
\bibitem [{\citenamefont {Henke}\ and\ \citenamefont
  {Papathanasiou}(2021)}]{Henke:2021ity}%
  \BibitemOpen
  \bibfield  {author} {\bibinfo {author} {\bibfnamefont {N.}~\bibnamefont
  {Henke}}\ and\ \bibinfo {author} {\bibfnamefont {G.}~\bibnamefont
  {Papathanasiou}},\ }\bibfield  {title} {\bibinfo {title} {{Singularities of
  eight- and nine-particle amplitudes from cluster algebras and tropical
  geometry}},\ }\href {https://doi.org/10.1007/JHEP10(2021)007} {\bibfield
  {journal} {\bibinfo  {journal} {JHEP}\ }\textbf {\bibinfo {volume} {10}},\
  \bibinfo {pages} {007}},\ \Eprint {https://arxiv.org/abs/2106.01392}
  {arXiv:2106.01392 [hep-th]} \BibitemShut {NoStop}%
\bibitem [{\citenamefont {Ren}\ \emph {et~al.}(2021)\citenamefont {Ren},
  \citenamefont {Spradlin},\ and\ \citenamefont {Volovich}}]{Ren:2021ztg}%
  \BibitemOpen
  \bibfield  {author} {\bibinfo {author} {\bibfnamefont {L.}~\bibnamefont
  {Ren}}, \bibinfo {author} {\bibfnamefont {M.}~\bibnamefont {Spradlin}},\ and\
  \bibinfo {author} {\bibfnamefont {A.}~\bibnamefont {Volovich}},\ }\bibfield
  {title} {\bibinfo {title} {{Symbol alphabets from tensor diagrams}},\ }\href
  {https://doi.org/10.1007/JHEP12(2021)079} {\bibfield  {journal} {\bibinfo
  {journal} {JHEP}\ }\textbf {\bibinfo {volume} {12}},\ \bibinfo {pages}
  {079}},\ \Eprint {https://arxiv.org/abs/2106.01405} {arXiv:2106.01405
  [hep-th]} \BibitemShut {NoStop}%
\bibitem [{\citenamefont {Yang}(2022)}]{Yang:2022gko}%
  \BibitemOpen
  \bibfield  {author} {\bibinfo {author} {\bibfnamefont {Q.}~\bibnamefont
  {Yang}},\ }\bibfield  {title} {\bibinfo {title} {{Schubert problems,
  positivity and symbol letters}},\ }\href
  {https://doi.org/10.1007/JHEP08(2022)168} {\bibfield  {journal} {\bibinfo
  {journal} {JHEP}\ }\textbf {\bibinfo {volume} {08}},\ \bibinfo {pages}
  {168}},\ \Eprint {https://arxiv.org/abs/2203.16112} {arXiv:2203.16112
  [hep-th]} \BibitemShut {NoStop}%
\bibitem [{\citenamefont {Bourjaily}\ \emph {et~al.}(2020)\citenamefont
  {Bourjaily}, \citenamefont {Gardi}, \citenamefont {McLeod},\ and\
  \citenamefont {Vergu}}]{Bourjaily:2019exo}%
  \BibitemOpen
  \bibfield  {author} {\bibinfo {author} {\bibfnamefont {J.~L.}\ \bibnamefont
  {Bourjaily}}, \bibinfo {author} {\bibfnamefont {E.}~\bibnamefont {Gardi}},
  \bibinfo {author} {\bibfnamefont {A.~J.}\ \bibnamefont {McLeod}},\ and\
  \bibinfo {author} {\bibfnamefont {C.}~\bibnamefont {Vergu}},\ }\bibfield
  {title} {\bibinfo {title} {{All-mass $n$-gon integrals in $n$ dimensions}},\
  }\href {https://doi.org/10.1007/JHEP08(2020)029} {\bibfield  {journal}
  {\bibinfo  {journal} {JHEP}\ }\textbf {\bibinfo {volume} {08}}\bibfield
  {number} {\bibinfo  {number} { (08)},\ \bibinfo {pages} {029}},\ }\Eprint
  {https://arxiv.org/abs/1912.11067} {arXiv:1912.11067 [hep-th]} \BibitemShut
  {NoStop}%
\bibitem [{\citenamefont {Dixon}\ \emph {et~al.}(2021)\citenamefont {Dixon},
  \citenamefont {McLeod},\ and\ \citenamefont {Wilhelm}}]{Dixon:2020bbt}%
  \BibitemOpen
  \bibfield  {author} {\bibinfo {author} {\bibfnamefont {L.~J.}\ \bibnamefont
  {Dixon}}, \bibinfo {author} {\bibfnamefont {A.~J.}\ \bibnamefont {McLeod}},\
  and\ \bibinfo {author} {\bibfnamefont {M.}~\bibnamefont {Wilhelm}},\
  }\bibfield  {title} {\bibinfo {title} {{A Three-Point Form Factor Through
  Five Loops}},\ }\href {https://doi.org/10.1007/JHEP04(2021)147} {\bibfield
  {journal} {\bibinfo  {journal} {JHEP}\ }\textbf {\bibinfo {volume} {04}},\
  \bibinfo {pages} {147}},\ \Eprint {https://arxiv.org/abs/2012.12286}
  {arXiv:2012.12286 [hep-th]} \BibitemShut {NoStop}%
\bibitem [{\citenamefont {He}\ \emph {et~al.}(2021{\natexlab{b}})\citenamefont
  {He}, \citenamefont {Li}, \citenamefont {Yang},\ and\ \citenamefont
  {Zhang}}]{He:2020lcu}%
  \BibitemOpen
  \bibfield  {author} {\bibinfo {author} {\bibfnamefont {S.}~\bibnamefont
  {He}}, \bibinfo {author} {\bibfnamefont {Z.}~\bibnamefont {Li}}, \bibinfo
  {author} {\bibfnamefont {Q.}~\bibnamefont {Yang}},\ and\ \bibinfo {author}
  {\bibfnamefont {C.}~\bibnamefont {Zhang}},\ }\bibfield  {title} {\bibinfo
  {title} {{Feynman Integrals and Scattering Amplitudes from Wilson Loops}},\
  }\href {https://doi.org/10.1103/PhysRevLett.126.231601} {\bibfield  {journal}
  {\bibinfo  {journal} {Phys. Rev. Lett.}\ }\textbf {\bibinfo {volume} {126}},\
  \bibinfo {pages} {231601} (\bibinfo {year} {2021}{\natexlab{b}})},\ \Eprint
  {https://arxiv.org/abs/2012.15042} {arXiv:2012.15042 [hep-th]} \BibitemShut
  {NoStop}%
\bibitem [{\citenamefont {He}\ \emph {et~al.}(2021{\natexlab{c}})\citenamefont
  {He}, \citenamefont {Li},\ and\ \citenamefont {Yang}}]{He:2021eec}%
  \BibitemOpen
  \bibfield  {author} {\bibinfo {author} {\bibfnamefont {S.}~\bibnamefont
  {He}}, \bibinfo {author} {\bibfnamefont {Z.}~\bibnamefont {Li}},\ and\
  \bibinfo {author} {\bibfnamefont {Q.}~\bibnamefont {Yang}},\ }\bibfield
  {title} {\bibinfo {title} {{Kinematics, cluster algebras and Feynman
  integrals}},\ }\href@noop {} {\  (\bibinfo {year} {2021}{\natexlab{c}})},\
  \Eprint {https://arxiv.org/abs/2112.11842} {arXiv:2112.11842 [hep-th]}
  \BibitemShut {NoStop}%
\bibitem [{\citenamefont {Abreu}\ \emph {et~al.}(2022)\citenamefont {Abreu},
  \citenamefont {Ita}, \citenamefont {Page},\ and\ \citenamefont
  {Tschernow}}]{Abreu:2021smk}%
  \BibitemOpen
  \bibfield  {author} {\bibinfo {author} {\bibfnamefont {S.}~\bibnamefont
  {Abreu}}, \bibinfo {author} {\bibfnamefont {H.}~\bibnamefont {Ita}}, \bibinfo
  {author} {\bibfnamefont {B.}~\bibnamefont {Page}},\ and\ \bibinfo {author}
  {\bibfnamefont {W.}~\bibnamefont {Tschernow}},\ }\bibfield  {title} {\bibinfo
  {title} {{Two-loop hexa-box integrals for non-planar five-point one-mass
  processes}},\ }\href {https://doi.org/10.1007/JHEP03(2022)182} {\bibfield
  {journal} {\bibinfo  {journal} {JHEP}\ }\textbf {\bibinfo {volume} {03}},\
  \bibinfo {pages} {182}},\ \Eprint {https://arxiv.org/abs/2107.14180}
  {arXiv:2107.14180 [hep-ph]} \BibitemShut {NoStop}%
\bibitem [{\citenamefont {He}\ \emph {et~al.}(2022{\natexlab{a}})\citenamefont
  {He}, \citenamefont {Liu}, \citenamefont {Tang},\ and\ \citenamefont
  {Yang}}]{He:2022tph}%
  \BibitemOpen
  \bibfield  {author} {\bibinfo {author} {\bibfnamefont {S.}~\bibnamefont
  {He}}, \bibinfo {author} {\bibfnamefont {J.}~\bibnamefont {Liu}}, \bibinfo
  {author} {\bibfnamefont {Y.}~\bibnamefont {Tang}},\ and\ \bibinfo {author}
  {\bibfnamefont {Q.}~\bibnamefont {Yang}},\ }\bibfield  {title} {\bibinfo
  {title} {{The symbology of Feynman integrals from twistor geometries}},\
  }\href@noop {} {\  (\bibinfo {year} {2022}{\natexlab{a}})},\ \Eprint
  {https://arxiv.org/abs/2207.13482} {arXiv:2207.13482 [hep-th]} \BibitemShut
  {NoStop}%
\bibitem [{\citenamefont {Henn}\ \emph {et~al.}(2023)\citenamefont {Henn},
  \citenamefont {Matija\v{s}i\'c},\ and\ \citenamefont
  {Miczajka}}]{Henn:2022ydo}%
  \BibitemOpen
  \bibfield  {author} {\bibinfo {author} {\bibfnamefont {J.~M.}\ \bibnamefont
  {Henn}}, \bibinfo {author} {\bibfnamefont {A.}~\bibnamefont
  {Matija\v{s}i\'c}},\ and\ \bibinfo {author} {\bibfnamefont {J.}~\bibnamefont
  {Miczajka}},\ }\bibfield  {title} {\bibinfo {title} {{One-loop hexagon
  integral to higher orders in the dimensional regulator}},\ }\href
  {https://doi.org/10.1007/JHEP01(2023)096} {\bibfield  {journal} {\bibinfo
  {journal} {JHEP}\ }\textbf {\bibinfo {volume} {01}},\ \bibinfo {pages}
  {096}},\ \Eprint {https://arxiv.org/abs/2210.13505} {arXiv:2210.13505
  [hep-th]} \BibitemShut {NoStop}%
\bibitem [{\citenamefont {Chen}(1977)}]{Chen}%
  \BibitemOpen
  \bibfield  {author} {\bibinfo {author} {\bibfnamefont {K.-T.}\ \bibnamefont
  {Chen}},\ }\bibfield  {title} {\bibinfo {title} {Iterated path integrals},\
  }\href {http://projecteuclid.org/euclid.bams/1183539443} {\bibfield
  {journal} {\bibinfo  {journal} {Bull. Amer. Math. Soc.}\ }\textbf {\bibinfo
  {volume} {83}},\ \bibinfo {pages} {831} (\bibinfo {year} {1977})}\BibitemShut
  {NoStop}%
\bibitem [{\citenamefont {Goncharov}(1995)}]{G91b}%
  \BibitemOpen
  \bibfield  {author} {\bibinfo {author} {\bibfnamefont {A.~B.}\ \bibnamefont
  {Goncharov}},\ }\bibfield  {title} {\bibinfo {title} {Geometry of
  configurations, polylogarithms, and motivic cohomology},\ }\href
  {https://doi.org/http://dx.doi.org/10.1006/aima.1995.1045} {\bibfield
  {journal} {\bibinfo  {journal} {Adv. Math.}\ }\textbf {\bibinfo {volume}
  {114}},\ \bibinfo {pages} {197} (\bibinfo {year} {1995})}\BibitemShut
  {NoStop}%
\bibitem [{\citenamefont {Goncharov}(1998)}]{Goncharov:1998kja}%
  \BibitemOpen
  \bibfield  {author} {\bibinfo {author} {\bibfnamefont {A.~B.}\ \bibnamefont
  {Goncharov}},\ }\bibfield  {title} {\bibinfo {title} {{Multiple
  polylogarithms, cyclotomy and modular complexes}},\ }\href
  {https://doi.org/10.4310/MRL.1998.v5.n4.a7} {\bibfield  {journal} {\bibinfo
  {journal} {Math. Res. Lett.}\ }\textbf {\bibinfo {volume} {5}},\ \bibinfo
  {pages} {497} (\bibinfo {year} {1998})},\ \Eprint
  {https://arxiv.org/abs/1105.2076} {arXiv:1105.2076 [math.AG]} \BibitemShut
  {NoStop}%
\bibitem [{\citenamefont {Remiddi}\ and\ \citenamefont
  {Vermaseren}(2000)}]{Remiddi:1999ew}%
  \BibitemOpen
  \bibfield  {author} {\bibinfo {author} {\bibfnamefont {E.}~\bibnamefont
  {Remiddi}}\ and\ \bibinfo {author} {\bibfnamefont {J.}~\bibnamefont
  {Vermaseren}},\ }\bibfield  {title} {\bibinfo {title} {{Harmonic
  polylogarithms}},\ }\href {https://doi.org/10.1142/S0217751X00000367}
  {\bibfield  {journal} {\bibinfo  {journal} {Int. J. Mod. Phys. A}\ }\textbf
  {\bibinfo {volume} {15}},\ \bibinfo {pages} {725} (\bibinfo {year} {2000})},\
  \Eprint {https://arxiv.org/abs/hep-ph/9905237} {arXiv:hep-ph/9905237}
  \BibitemShut {NoStop}%
\bibitem [{\citenamefont {Borwein}\ \emph {et~al.}(2001)\citenamefont
  {Borwein}, \citenamefont {Bradley}, \citenamefont {Broadhurst},\ and\
  \citenamefont {Lisonek}}]{Borwein:1999js}%
  \BibitemOpen
  \bibfield  {author} {\bibinfo {author} {\bibfnamefont {J.~M.}\ \bibnamefont
  {Borwein}}, \bibinfo {author} {\bibfnamefont {D.~M.}\ \bibnamefont
  {Bradley}}, \bibinfo {author} {\bibfnamefont {D.~J.}\ \bibnamefont
  {Broadhurst}},\ and\ \bibinfo {author} {\bibfnamefont {P.}~\bibnamefont
  {Lisonek}},\ }\bibfield  {title} {\bibinfo {title} {{Special values of
  multiple polylogarithms}},\ }\href
  {https://doi.org/10.1090/S0002-9947-00-02616-7} {\bibfield  {journal}
  {\bibinfo  {journal} {Trans. Am. Math. Soc.}\ }\textbf {\bibinfo {volume}
  {353}},\ \bibinfo {pages} {907} (\bibinfo {year} {2001})},\ \Eprint
  {https://arxiv.org/abs/math/9910045} {arXiv:math/9910045 [math-ca]}
  \BibitemShut {NoStop}%
\bibitem [{\citenamefont {Moch}\ \emph {et~al.}(2002)\citenamefont {Moch},
  \citenamefont {Uwer},\ and\ \citenamefont {Weinzierl}}]{Moch:2001zr}%
  \BibitemOpen
  \bibfield  {author} {\bibinfo {author} {\bibfnamefont {S.}~\bibnamefont
  {Moch}}, \bibinfo {author} {\bibfnamefont {P.}~\bibnamefont {Uwer}},\ and\
  \bibinfo {author} {\bibfnamefont {S.}~\bibnamefont {Weinzierl}},\ }\bibfield
  {title} {\bibinfo {title} {{Nested sums, expansion of transcendental
  functions and multiscale multiloop integrals}},\ }\href
  {https://doi.org/10.1063/1.1471366} {\bibfield  {journal} {\bibinfo
  {journal} {J. Math. Phys.}\ }\textbf {\bibinfo {volume} {43}},\ \bibinfo
  {pages} {3363} (\bibinfo {year} {2002})},\ \Eprint
  {https://arxiv.org/abs/hep-ph/0110083} {arXiv:hep-ph/0110083 [hep-ph]}
  \BibitemShut {NoStop}%
\bibitem [{\citenamefont {Goncharov}(2005)}]{Gonch2}%
  \BibitemOpen
  \bibfield  {author} {\bibinfo {author} {\bibfnamefont {A.~B.}\ \bibnamefont
  {Goncharov}},\ }\bibfield  {title} {\bibinfo {title} {Galois symmetries of
  fundamental groupoids and noncommutative geometry},\ }\href
  {https://doi.org/10.1215/S0012-7094-04-12822-2} {\bibfield  {journal}
  {\bibinfo  {journal} {Duke Math. J.}\ }\textbf {\bibinfo {volume} {128}},\
  \bibinfo {pages} {209} (\bibinfo {year} {2005})},\ \Eprint
  {https://arxiv.org/abs/math/0208144} {arXiv:math/0208144 [math.AG]}
  \BibitemShut {NoStop}%
\bibitem [{\citenamefont {Goncharov}\ \emph {et~al.}(2010)\citenamefont
  {Goncharov}, \citenamefont {Spradlin}, \citenamefont {Vergu},\ and\
  \citenamefont {Volovich}}]{Goncharov:2010jf}%
  \BibitemOpen
  \bibfield  {author} {\bibinfo {author} {\bibfnamefont {A.~B.}\ \bibnamefont
  {Goncharov}}, \bibinfo {author} {\bibfnamefont {M.}~\bibnamefont {Spradlin}},
  \bibinfo {author} {\bibfnamefont {C.}~\bibnamefont {Vergu}},\ and\ \bibinfo
  {author} {\bibfnamefont {A.}~\bibnamefont {Volovich}},\ }\bibfield  {title}
  {\bibinfo {title} {{Classical Polylogarithms for Amplitudes and Wilson
  Loops}},\ }\href {https://doi.org/10.1103/PhysRevLett.105.151605} {\bibfield
  {journal} {\bibinfo  {journal} {Phys. Rev. Lett.}\ }\textbf {\bibinfo
  {volume} {105}},\ \bibinfo {pages} {151605} (\bibinfo {year} {2010})},\
  \Eprint {https://arxiv.org/abs/1006.5703} {arXiv:1006.5703 [hep-th]}
  \BibitemShut {NoStop}%
\bibitem [{\citenamefont {Brown}(2012{\natexlab{a}})}]{Brown:2011ik}%
  \BibitemOpen
  \bibfield  {author} {\bibinfo {author} {\bibfnamefont {F.}~\bibnamefont
  {Brown}},\ }\bibfield  {title} {\bibinfo {title} {{On the decomposition of
  motivic multiple zeta values}},\ }\href@noop {} {\bibfield  {journal}
  {\bibinfo  {journal} {{Adv. Studies in Pure Math.}}\ }\textbf {\bibinfo
  {volume} {63}},\ \bibinfo {pages} {31} (\bibinfo {year}
  {2012}{\natexlab{a}})},\ \Eprint {https://arxiv.org/abs/1102.1310}
  {arXiv:1102.1310 [math.NT]} \BibitemShut {NoStop}%
\bibitem [{\citenamefont {Brown}(2012{\natexlab{b}})}]{Brown1102.1312}%
  \BibitemOpen
  \bibfield  {author} {\bibinfo {author} {\bibfnamefont {F.}~\bibnamefont
  {Brown}},\ }\bibfield  {title} {\bibinfo {title} {Mixed {T}ate motives over
  {$\mathbb{Z}$}},\ }\href {https://doi.org/10.4007/annals.2012.175.2.10}
  {\bibfield  {journal} {\bibinfo  {journal} {Ann. of Math. (2)}\ }\textbf
  {\bibinfo {volume} {175}},\ \bibinfo {pages} {949} (\bibinfo {year}
  {2012}{\natexlab{b}})},\ \Eprint {https://arxiv.org/abs/1102.1312}
  {arXiv:1102.1312 [math.AG]} \BibitemShut {NoStop}%
\bibitem [{\citenamefont {Duhr}\ \emph {et~al.}(2012)\citenamefont {Duhr},
  \citenamefont {Gangl},\ and\ \citenamefont {Rhodes}}]{Duhr:2011zq}%
  \BibitemOpen
  \bibfield  {author} {\bibinfo {author} {\bibfnamefont {C.}~\bibnamefont
  {Duhr}}, \bibinfo {author} {\bibfnamefont {H.}~\bibnamefont {Gangl}},\ and\
  \bibinfo {author} {\bibfnamefont {J.~R.}\ \bibnamefont {Rhodes}},\ }\bibfield
   {title} {\bibinfo {title} {{From polygons and symbols to polylogarithmic
  functions}},\ }\href {https://doi.org/10.1007/JHEP10(2012)075} {\bibfield
  {journal} {\bibinfo  {journal} {JHEP}\ }\textbf {\bibinfo {volume} {10}},\
  \bibinfo {pages} {075}},\ \Eprint {https://arxiv.org/abs/1110.0458}
  {arXiv:1110.0458 [math-ph]} \BibitemShut {NoStop}%
\bibitem [{\citenamefont {Duhr}(2012)}]{Duhr:2012fh}%
  \BibitemOpen
  \bibfield  {author} {\bibinfo {author} {\bibfnamefont {C.}~\bibnamefont
  {Duhr}},\ }\bibfield  {title} {\bibinfo {title} {{Hopf algebras, coproducts
  and symbols: an application to Higgs boson amplitudes}},\ }\href
  {https://doi.org/10.1007/JHEP08(2012)043} {\bibfield  {journal} {\bibinfo
  {journal} {JHEP}\ }\textbf {\bibinfo {volume} {1208}},\ \bibinfo {pages}
  {043}},\ \Eprint {https://arxiv.org/abs/1203.0454} {arXiv:1203.0454 [hep-ph]}
  \BibitemShut {NoStop}%
\bibitem [{\citenamefont {Dixon}\ \emph {et~al.}(2011)\citenamefont {Dixon},
  \citenamefont {Drummond},\ and\ \citenamefont {Henn}}]{Dixon:2011pw}%
  \BibitemOpen
  \bibfield  {author} {\bibinfo {author} {\bibfnamefont {L.~J.}\ \bibnamefont
  {Dixon}}, \bibinfo {author} {\bibfnamefont {J.~M.}\ \bibnamefont
  {Drummond}},\ and\ \bibinfo {author} {\bibfnamefont {J.~M.}\ \bibnamefont
  {Henn}},\ }\bibfield  {title} {\bibinfo {title} {{Bootstrapping the
  three-loop hexagon}},\ }\href {https://doi.org/10.1007/JHEP11(2011)023}
  {\bibfield  {journal} {\bibinfo  {journal} {JHEP}\ }\textbf {\bibinfo
  {volume} {1111}},\ \bibinfo {pages} {023}},\ \Eprint
  {https://arxiv.org/abs/1108.4461} {arXiv:1108.4461 [hep-th]} \BibitemShut
  {NoStop}%
\bibitem [{\citenamefont {Dixon}\ \emph {et~al.}(2012)\citenamefont {Dixon},
  \citenamefont {Drummond},\ and\ \citenamefont {Henn}}]{Dixon:2011nj}%
  \BibitemOpen
  \bibfield  {author} {\bibinfo {author} {\bibfnamefont {L.~J.}\ \bibnamefont
  {Dixon}}, \bibinfo {author} {\bibfnamefont {J.~M.}\ \bibnamefont
  {Drummond}},\ and\ \bibinfo {author} {\bibfnamefont {J.~M.}\ \bibnamefont
  {Henn}},\ }\bibfield  {title} {\bibinfo {title} {{Analytic result for the
  two-loop six-point NMHV amplitude in $\mathcal{N}=4$ super Yang-Mills
  theory}},\ }\href {https://doi.org/10.1007/JHEP01(2012)024} {\bibfield
  {journal} {\bibinfo  {journal} {JHEP}\ }\textbf {\bibinfo {volume} {1201}},\
  \bibinfo {pages} {024}},\ \Eprint {https://arxiv.org/abs/1111.1704}
  {arXiv:1111.1704 [hep-th]} \BibitemShut {NoStop}%
\bibitem [{\citenamefont {Brandhuber}\ \emph {et~al.}(2012)\citenamefont
  {Brandhuber}, \citenamefont {Travaglini},\ and\ \citenamefont
  {Yang}}]{Brandhuber:2012vm}%
  \BibitemOpen
  \bibfield  {author} {\bibinfo {author} {\bibfnamefont {A.}~\bibnamefont
  {Brandhuber}}, \bibinfo {author} {\bibfnamefont {G.}~\bibnamefont
  {Travaglini}},\ and\ \bibinfo {author} {\bibfnamefont {G.}~\bibnamefont
  {Yang}},\ }\bibfield  {title} {\bibinfo {title} {{Analytic two-loop form
  factors in $\mathcal{N}=4$ SYM}},\ }\href
  {https://doi.org/10.1007/JHEP05(2012)082} {\bibfield  {journal} {\bibinfo
  {journal} {JHEP}\ }\textbf {\bibinfo {volume} {05}},\ \bibinfo {pages}
  {082}},\ \Eprint {https://arxiv.org/abs/1201.4170} {arXiv:1201.4170 [hep-th]}
  \BibitemShut {NoStop}%
\bibitem [{\citenamefont {Dixon}\ \emph {et~al.}(2013)\citenamefont {Dixon},
  \citenamefont {Drummond}, \citenamefont {von Hippel},\ and\ \citenamefont
  {Pennington}}]{Dixon:2013eka}%
  \BibitemOpen
  \bibfield  {author} {\bibinfo {author} {\bibfnamefont {L.~J.}\ \bibnamefont
  {Dixon}}, \bibinfo {author} {\bibfnamefont {J.~M.}\ \bibnamefont {Drummond}},
  \bibinfo {author} {\bibfnamefont {M.}~\bibnamefont {von Hippel}},\ and\
  \bibinfo {author} {\bibfnamefont {J.}~\bibnamefont {Pennington}},\ }\bibfield
   {title} {\bibinfo {title} {{Hexagon functions and the three-loop remainder
  function}},\ }\href {https://doi.org/10.1007/JHEP12(2013)049} {\bibfield
  {journal} {\bibinfo  {journal} {JHEP}\ }\textbf {\bibinfo {volume} {1312}},\
  \bibinfo {pages} {049}},\ \Eprint {https://arxiv.org/abs/1308.2276}
  {arXiv:1308.2276 [hep-th]} \BibitemShut {NoStop}%
\bibitem [{\citenamefont {Dixon}\ and\ \citenamefont {von
  Hippel}(2014)}]{Dixon:2014iba}%
  \BibitemOpen
  \bibfield  {author} {\bibinfo {author} {\bibfnamefont {L.~J.}\ \bibnamefont
  {Dixon}}\ and\ \bibinfo {author} {\bibfnamefont {M.}~\bibnamefont {von
  Hippel}},\ }\bibfield  {title} {\bibinfo {title} {{Bootstrapping an NMHV
  amplitude through three loops}},\ }\href
  {https://doi.org/10.1007/JHEP10(2014)065} {\bibfield  {journal} {\bibinfo
  {journal} {JHEP}\ }\textbf {\bibinfo {volume} {1410}},\ \bibinfo {pages}
  {65}},\ \Eprint {https://arxiv.org/abs/1408.1505} {arXiv:1408.1505 [hep-th]}
  \BibitemShut {NoStop}%
\bibitem [{\citenamefont {Dixon}\ \emph {et~al.}(2014)\citenamefont {Dixon},
  \citenamefont {Drummond}, \citenamefont {Duhr},\ and\ \citenamefont
  {Pennington}}]{Dixon:2014voa}%
  \BibitemOpen
  \bibfield  {author} {\bibinfo {author} {\bibfnamefont {L.~J.}\ \bibnamefont
  {Dixon}}, \bibinfo {author} {\bibfnamefont {J.~M.}\ \bibnamefont {Drummond}},
  \bibinfo {author} {\bibfnamefont {C.}~\bibnamefont {Duhr}},\ and\ \bibinfo
  {author} {\bibfnamefont {J.}~\bibnamefont {Pennington}},\ }\bibfield  {title}
  {\bibinfo {title} {{The four-loop remainder function and multi-Regge behavior
  at NNLLA in planar $\mathcal{N} = 4$ super-Yang-Mills theory}},\ }\href
  {https://doi.org/10.1007/JHEP06(2014)116} {\bibfield  {journal} {\bibinfo
  {journal} {JHEP}\ }\textbf {\bibinfo {volume} {1406}},\ \bibinfo {pages}
  {116}},\ \Eprint {https://arxiv.org/abs/1402.3300} {arXiv:1402.3300 [hep-th]}
  \BibitemShut {NoStop}%
\bibitem [{\citenamefont {Dixon}\ \emph {et~al.}(2016)\citenamefont {Dixon},
  \citenamefont {von Hippel},\ and\ \citenamefont {McLeod}}]{Dixon:2015iva}%
  \BibitemOpen
  \bibfield  {author} {\bibinfo {author} {\bibfnamefont {L.~J.}\ \bibnamefont
  {Dixon}}, \bibinfo {author} {\bibfnamefont {M.}~\bibnamefont {von Hippel}},\
  and\ \bibinfo {author} {\bibfnamefont {A.~J.}\ \bibnamefont {McLeod}},\
  }\bibfield  {title} {\bibinfo {title} {{The four-loop six-gluon NMHV ratio
  function}},\ }\href {https://doi.org/10.1007/JHEP01(2016)053} {\bibfield
  {journal} {\bibinfo  {journal} {JHEP}\ }\textbf {\bibinfo {volume} {01}},\
  \bibinfo {pages} {053}},\ \Eprint {https://arxiv.org/abs/1509.08127}
  {arXiv:1509.08127 [hep-th]} \BibitemShut {NoStop}%
\bibitem [{\citenamefont {Caron-Huot}\ \emph {et~al.}(2016)\citenamefont
  {Caron-Huot}, \citenamefont {Dixon}, \citenamefont {McLeod},\ and\
  \citenamefont {von Hippel}}]{Caron-Huot:2016owq}%
  \BibitemOpen
  \bibfield  {author} {\bibinfo {author} {\bibfnamefont {S.}~\bibnamefont
  {Caron-Huot}}, \bibinfo {author} {\bibfnamefont {L.~J.}\ \bibnamefont
  {Dixon}}, \bibinfo {author} {\bibfnamefont {A.}~\bibnamefont {McLeod}},\ and\
  \bibinfo {author} {\bibfnamefont {M.}~\bibnamefont {von Hippel}},\ }\bibfield
   {title} {\bibinfo {title} {{Bootstrapping a Five-Loop Amplitude Using
  Steinmann Relations}},\ }\href
  {https://doi.org/10.1103/PhysRevLett.117.241601} {\bibfield  {journal}
  {\bibinfo  {journal} {Phys. Rev. Lett.}\ }\textbf {\bibinfo {volume} {117}},\
  \bibinfo {pages} {241601} (\bibinfo {year} {2016})},\ \Eprint
  {https://arxiv.org/abs/1609.00669} {arXiv:1609.00669 [hep-th]} \BibitemShut
  {NoStop}%
\bibitem [{\citenamefont {Dixon}\ \emph {et~al.}(2017)\citenamefont {Dixon},
  \citenamefont {Drummond}, \citenamefont {Harrington}, \citenamefont {McLeod},
  \citenamefont {Papathanasiou},\ and\ \citenamefont
  {Spradlin}}]{Dixon:2016nkn}%
  \BibitemOpen
  \bibfield  {author} {\bibinfo {author} {\bibfnamefont {L.~J.}\ \bibnamefont
  {Dixon}}, \bibinfo {author} {\bibfnamefont {J.}~\bibnamefont {Drummond}},
  \bibinfo {author} {\bibfnamefont {T.}~\bibnamefont {Harrington}}, \bibinfo
  {author} {\bibfnamefont {A.~J.}\ \bibnamefont {McLeod}}, \bibinfo {author}
  {\bibfnamefont {G.}~\bibnamefont {Papathanasiou}},\ and\ \bibinfo {author}
  {\bibfnamefont {M.}~\bibnamefont {Spradlin}},\ }\bibfield  {title} {\bibinfo
  {title} {{Heptagons from the Steinmann Cluster Bootstrap}},\ }\href
  {https://doi.org/10.1007/JHEP02(2017)137} {\bibfield  {journal} {\bibinfo
  {journal} {JHEP}\ }\textbf {\bibinfo {volume} {02}},\ \bibinfo {pages}
  {137}},\ \Eprint {https://arxiv.org/abs/1612.08976} {arXiv:1612.08976
  [hep-th]} \BibitemShut {NoStop}%
\bibitem [{\citenamefont {Caron-Huot}\ \emph
  {et~al.}(2019{\natexlab{b}})\citenamefont {Caron-Huot}, \citenamefont
  {Dixon}, \citenamefont {Dulat}, \citenamefont {von Hippel}, \citenamefont
  {McLeod},\ and\ \citenamefont {Papathanasiou}}]{Caron-Huot:2019vjl}%
  \BibitemOpen
  \bibfield  {author} {\bibinfo {author} {\bibfnamefont {S.}~\bibnamefont
  {Caron-Huot}}, \bibinfo {author} {\bibfnamefont {L.~J.}\ \bibnamefont
  {Dixon}}, \bibinfo {author} {\bibfnamefont {F.}~\bibnamefont {Dulat}},
  \bibinfo {author} {\bibfnamefont {M.}~\bibnamefont {von Hippel}}, \bibinfo
  {author} {\bibfnamefont {A.~J.}\ \bibnamefont {McLeod}},\ and\ \bibinfo
  {author} {\bibfnamefont {G.}~\bibnamefont {Papathanasiou}},\ }\bibfield
  {title} {\bibinfo {title} {{Six-Gluon amplitudes in planar $ \mathcal{N} $ =
  4 super-Yang-Mills theory at six and seven loops}},\ }\href
  {https://doi.org/10.1007/JHEP08(2019)016} {\bibfield  {journal} {\bibinfo
  {journal} {JHEP}\ }\textbf {\bibinfo {volume} {08}},\ \bibinfo {pages}
  {016}},\ \Eprint {https://arxiv.org/abs/1903.10890} {arXiv:1903.10890
  [hep-th]} \BibitemShut {NoStop}%
\bibitem [{\citenamefont {Dixon}\ and\ \citenamefont
  {Liu}(2020)}]{Dixon:2020cnr}%
  \BibitemOpen
  \bibfield  {author} {\bibinfo {author} {\bibfnamefont {L.~J.}\ \bibnamefont
  {Dixon}}\ and\ \bibinfo {author} {\bibfnamefont {Y.-T.}\ \bibnamefont
  {Liu}},\ }\bibfield  {title} {\bibinfo {title} {{Lifting Heptagon Symbols to
  Functions}},\ }\href {https://doi.org/10.1007/JHEP10(2020)031} {\bibfield
  {journal} {\bibinfo  {journal} {JHEP}\ }\textbf {\bibinfo {volume} {10}},\
  \bibinfo {pages} {031}},\ \Eprint {https://arxiv.org/abs/2007.12966}
  {arXiv:2007.12966 [hep-th]} \BibitemShut {NoStop}%
\bibitem [{\citenamefont {Dixon}\ \emph
  {et~al.}(2022{\natexlab{a}})\citenamefont {Dixon}, \citenamefont
  {G{\"u}rdo{\u{g}}an}, \citenamefont {McLeod},\ and\ \citenamefont
  {Wilhelm}}]{Dixon:2022rse}%
  \BibitemOpen
  \bibfield  {author} {\bibinfo {author} {\bibfnamefont {L.~J.}\ \bibnamefont
  {Dixon}}, \bibinfo {author} {\bibfnamefont {{\"O}.}~\bibnamefont
  {G{\"u}rdo{\u{g}}an}}, \bibinfo {author} {\bibfnamefont {A.~J.}\ \bibnamefont
  {McLeod}},\ and\ \bibinfo {author} {\bibfnamefont {M.}~\bibnamefont
  {Wilhelm}},\ }\bibfield  {title} {\bibinfo {title} {{Bootstrapping a
  stress-tensor form factor through eight loops}},\ }\href
  {https://doi.org/10.1007/JHEP07(2022)153} {\bibfield  {journal} {\bibinfo
  {journal} {JHEP}\ }\textbf {\bibinfo {volume} {07}},\ \bibinfo {pages}
  {153}},\ \Eprint {https://arxiv.org/abs/2204.11901} {arXiv:2204.11901
  [hep-th]} \BibitemShut {NoStop}%
\bibitem [{\citenamefont {Dixon}\ \emph
  {et~al.}(2022{\natexlab{b}})\citenamefont {Dixon}, \citenamefont
  {G{\"u}rdo{\u{g}}an}, \citenamefont {McLeod},\ and\ \citenamefont
  {Wilhelm}}]{Dixon:2021tdw}%
  \BibitemOpen
  \bibfield  {author} {\bibinfo {author} {\bibfnamefont {L.~J.}\ \bibnamefont
  {Dixon}}, \bibinfo {author} {\bibfnamefont {{\"O}.}~\bibnamefont
  {G{\"u}rdo{\u{g}}an}}, \bibinfo {author} {\bibfnamefont {A.~J.}\ \bibnamefont
  {McLeod}},\ and\ \bibinfo {author} {\bibfnamefont {M.}~\bibnamefont
  {Wilhelm}},\ }\bibfield  {title} {\bibinfo {title} {{Folding Amplitudes into
  Form Factors: An Antipodal Duality}},\ }\href
  {https://doi.org/10.1103/PhysRevLett.128.111602} {\bibfield  {journal}
  {\bibinfo  {journal} {Phys. Rev. Lett.}\ }\textbf {\bibinfo {volume} {128}},\
  \bibinfo {pages} {111602} (\bibinfo {year} {2022}{\natexlab{b}})},\ \Eprint
  {https://arxiv.org/abs/2112.06243} {arXiv:2112.06243 [hep-th]} \BibitemShut
  {NoStop}%
\bibitem [{\citenamefont {Dixon}\ and\ \citenamefont
  {Liu}()}]{ToAppearAmpEightLoop}%
  \BibitemOpen
  \bibfield  {author} {\bibinfo {author} {\bibfnamefont {L.~J.}\ \bibnamefont
  {Dixon}}\ and\ \bibinfo {author} {\bibfnamefont {Y.-T.}\ \bibnamefont
  {Liu}},\ }\href@noop {} {\bibinfo {title} {{An Eight Loop Amplitude via
  Antipodal Duality}}},\ \bibinfo {howpublished} {to appear}\BibitemShut
  {NoStop}%
\bibitem [{\citenamefont {Liu}(2022)}]{Liu:2022vck}%
  \BibitemOpen
  \bibfield  {author} {\bibinfo {author} {\bibfnamefont {Y.-T.}\ \bibnamefont
  {Liu}},\ }\bibfield  {title} {\bibinfo {title} {{Antipodal symmetry of
  two-loop MHV amplitudes}},\ }\href {https://doi.org/10.1007/JHEP09(2022)131}
  {\bibfield  {journal} {\bibinfo  {journal} {JHEP}\ }\textbf {\bibinfo
  {volume} {09}},\ \bibinfo {pages} {131}},\ \Eprint
  {https://arxiv.org/abs/2207.11815} {arXiv:2207.11815 [hep-th]} \BibitemShut
  {NoStop}%
\bibitem [{\citenamefont {Abreu}\ \emph {et~al.}(2020)\citenamefont {Abreu},
  \citenamefont {Ita}, \citenamefont {Moriello}, \citenamefont {Page},
  \citenamefont {Tschernow},\ and\ \citenamefont {Zeng}}]{Abreu:2020jxa}%
  \BibitemOpen
  \bibfield  {author} {\bibinfo {author} {\bibfnamefont {S.}~\bibnamefont
  {Abreu}}, \bibinfo {author} {\bibfnamefont {H.}~\bibnamefont {Ita}}, \bibinfo
  {author} {\bibfnamefont {F.}~\bibnamefont {Moriello}}, \bibinfo {author}
  {\bibfnamefont {B.}~\bibnamefont {Page}}, \bibinfo {author} {\bibfnamefont
  {W.}~\bibnamefont {Tschernow}},\ and\ \bibinfo {author} {\bibfnamefont
  {M.}~\bibnamefont {Zeng}},\ }\bibfield  {title} {\bibinfo {title} {{Two-Loop
  Integrals for Planar Five-Point One-Mass Processes}},\ }\href
  {https://doi.org/10.1007/JHEP11(2020)117} {\bibfield  {journal} {\bibinfo
  {journal} {JHEP}\ }\textbf {\bibinfo {volume} {11}},\ \bibinfo {pages}
  {117}},\ \Eprint {https://arxiv.org/abs/2005.04195} {arXiv:2005.04195
  [hep-ph]} \BibitemShut {NoStop}%
\bibitem [{\citenamefont {Abreu}\ \emph {et~al.}()\citenamefont {Abreu},
  \citenamefont {Chicherin}, \citenamefont {Ita}, \citenamefont {Page},
  \citenamefont {Sotnikov}, \citenamefont {Tschernow},\ and\ \citenamefont
  {Zoia}}]{ToAppearAbreuEtal}%
  \BibitemOpen
  \bibfield  {author} {\bibinfo {author} {\bibfnamefont {S.}~\bibnamefont
  {Abreu}}, \bibinfo {author} {\bibfnamefont {D.}~\bibnamefont {Chicherin}},
  \bibinfo {author} {\bibfnamefont {H.}~\bibnamefont {Ita}}, \bibinfo {author}
  {\bibfnamefont {B.}~\bibnamefont {Page}}, \bibinfo {author} {\bibfnamefont
  {V.}~\bibnamefont {Sotnikov}}, \bibinfo {author} {\bibfnamefont
  {W.}~\bibnamefont {Tschernow}},\ and\ \bibinfo {author} {\bibfnamefont
  {S.}~\bibnamefont {Zoia}},\ }\href@noop {} {}\bibinfo {howpublished} {to
  appear}\BibitemShut {NoStop}%
\bibitem [{\citenamefont {Abreu}()}]{samuel_private}%
  \BibitemOpen
  \bibfield  {author} {\bibinfo {author} {\bibfnamefont {S.}~\bibnamefont
  {Abreu}},\ }\href@noop {} {}\bibinfo {howpublished} {private
  communication}\BibitemShut {NoStop}%
\bibitem [{\citenamefont {Drummond}\ \emph {et~al.}(2015)\citenamefont
  {Drummond}, \citenamefont {Papathanasiou},\ and\ \citenamefont
  {Spradlin}}]{Drummond:2014ffa}%
  \BibitemOpen
  \bibfield  {author} {\bibinfo {author} {\bibfnamefont {J.~M.}\ \bibnamefont
  {Drummond}}, \bibinfo {author} {\bibfnamefont {G.}~\bibnamefont
  {Papathanasiou}},\ and\ \bibinfo {author} {\bibfnamefont {M.}~\bibnamefont
  {Spradlin}},\ }\bibfield  {title} {\bibinfo {title} {{A Symbol of Uniqueness:
  The Cluster Bootstrap for the 3-Loop MHV Heptagon}},\ }\href
  {https://doi.org/10.1007/JHEP03(2015)072} {\bibfield  {journal} {\bibinfo
  {journal} {JHEP}\ }\textbf {\bibinfo {volume} {03}},\ \bibinfo {pages}
  {072}},\ \Eprint {https://arxiv.org/abs/1412.3763} {arXiv:1412.3763 [hep-th]}
  \BibitemShut {NoStop}%
\bibitem [{\citenamefont {Guo}\ \emph {et~al.}(2022{\natexlab{a}})\citenamefont
  {Guo}, \citenamefont {Wang},\ and\ \citenamefont {Yang}}]{Guo:2022qgv}%
  \BibitemOpen
  \bibfield  {author} {\bibinfo {author} {\bibfnamefont {Y.}~\bibnamefont
  {Guo}}, \bibinfo {author} {\bibfnamefont {L.}~\bibnamefont {Wang}},\ and\
  \bibinfo {author} {\bibfnamefont {G.}~\bibnamefont {Yang}},\ }\bibfield
  {title} {\bibinfo {title} {{Analytic Four-Point Lightlike Form Factors and
  OPE of Null-Wrapped Polygons}},\ }\href@noop {} {\  (\bibinfo {year}
  {2022}{\natexlab{a}})},\ \Eprint {https://arxiv.org/abs/2209.06816}
  {arXiv:2209.06816 [hep-th]} \BibitemShut {NoStop}%
\bibitem [{\citenamefont {Brandhuber}\ \emph
  {et~al.}(2011{\natexlab{b}})\citenamefont {Brandhuber}, \citenamefont
  {G{\"u}rdo{\u{g}}an}, \citenamefont {Mooney}, \citenamefont {Travaglini},\
  and\ \citenamefont {Yang}}]{Brandhuber:2011tv}%
  \BibitemOpen
  \bibfield  {author} {\bibinfo {author} {\bibfnamefont {A.}~\bibnamefont
  {Brandhuber}}, \bibinfo {author} {\bibfnamefont {{\"O}.}~\bibnamefont
  {G{\"u}rdo{\u{g}}an}}, \bibinfo {author} {\bibfnamefont {R.}~\bibnamefont
  {Mooney}}, \bibinfo {author} {\bibfnamefont {G.}~\bibnamefont {Travaglini}},\
  and\ \bibinfo {author} {\bibfnamefont {G.}~\bibnamefont {Yang}},\ }\bibfield
  {title} {\bibinfo {title} {{Harmony of Super Form Factors}},\ }\href
  {https://doi.org/10.1007/JHEP10(2011)046} {\bibfield  {journal} {\bibinfo
  {journal} {JHEP}\ }\textbf {\bibinfo {volume} {10}},\ \bibinfo {pages}
  {046}},\ \Eprint {https://arxiv.org/abs/1107.5067} {arXiv:1107.5067 [hep-th]}
  \BibitemShut {NoStop}%
\bibitem [{\citenamefont {Duhr}(2015)}]{Duhr:2014woa}%
  \BibitemOpen
  \bibfield  {author} {\bibinfo {author} {\bibfnamefont {C.}~\bibnamefont
  {Duhr}},\ }\bibfield  {title} {\bibinfo {title} {{Mathematical aspects of
  scattering amplitudes}},\ }in\ \href
  {https://doi.org/10.1142/9789814678766_0010} {\emph {\bibinfo {booktitle}
  {{Theoretical Advanced Study Institute in Elementary Particle Physics}:
  {Journeys Through the Precision Frontier: Amplitudes for Colliders}}}}\
  (\bibinfo {year} {2015})\ pp.\ \bibinfo {pages} {419--476},\ \Eprint
  {https://arxiv.org/abs/1411.7538} {arXiv:1411.7538 [hep-ph]} \BibitemShut
  {NoStop}%
\bibitem [{\citenamefont {Gaiotto}\ \emph {et~al.}(2011)\citenamefont
  {Gaiotto}, \citenamefont {Maldacena}, \citenamefont {Sever},\ and\
  \citenamefont {Vieira}}]{Gaiotto:2011dt}%
  \BibitemOpen
  \bibfield  {author} {\bibinfo {author} {\bibfnamefont {D.}~\bibnamefont
  {Gaiotto}}, \bibinfo {author} {\bibfnamefont {J.}~\bibnamefont {Maldacena}},
  \bibinfo {author} {\bibfnamefont {A.}~\bibnamefont {Sever}},\ and\ \bibinfo
  {author} {\bibfnamefont {P.}~\bibnamefont {Vieira}},\ }\bibfield  {title}
  {\bibinfo {title} {{Pulling the straps of polygons}},\ }\href
  {https://doi.org/10.1007/JHEP12(2011)011} {\bibfield  {journal} {\bibinfo
  {journal} {JHEP}\ }\textbf {\bibinfo {volume} {12}},\ \bibinfo {pages}
  {011}},\ \Eprint {https://arxiv.org/abs/1102.0062} {arXiv:1102.0062 [hep-th]}
  \BibitemShut {NoStop}%
\bibitem [{\citenamefont {{The SpaSM group}}(2017)}]{spasm}%
  \BibitemOpen
  \bibfield  {author} {\bibinfo {author} {\bibnamefont {{The SpaSM group}}},\
  }\href@noop {} {\emph {\bibinfo {title} {{SpaSM}: a Sparse direct Solver
  Modulo $p$}}},\ \bibinfo {edition} {v1.2}\ ed. (\bibinfo {year} {2017}),\
  \bibinfo {note} {\url{http://github.com/cbouilla/spasm}}\BibitemShut
  {NoStop}%
\bibitem [{\citenamefont {Steinmann}(1960{\natexlab{a}})}]{Steinmann}%
  \BibitemOpen
  \bibfield  {author} {\bibinfo {author} {\bibfnamefont {O.}~\bibnamefont
  {Steinmann}},\ }\bibfield  {title} {\bibinfo {title} {{\"Uber den
  Zusammenhang zwischen den Wightmanfunktionen und der retardierten
  Kommutatoren}},\ }\href@noop {} {\bibfield  {journal} {\bibinfo  {journal}
  {Helv. Physica Acta}\ }\textbf {\bibinfo {volume} {33}},\ \bibinfo {pages}
  {257} (\bibinfo {year} {1960}{\natexlab{a}})}\BibitemShut {NoStop}%
\bibitem [{\citenamefont {Steinmann}(1960{\natexlab{b}})}]{Steinmann2}%
  \BibitemOpen
  \bibfield  {author} {\bibinfo {author} {\bibfnamefont {O.}~\bibnamefont
  {Steinmann}},\ }\bibfield  {title} {\bibinfo {title} {{Wightman-Funktionen
  und retardierten Kommutatoren. II}},\ }\href@noop {} {\bibfield  {journal}
  {\bibinfo  {journal} {Helv. Physica Acta}\ }\textbf {\bibinfo {volume}
  {33}},\ \bibinfo {pages} {347} (\bibinfo {year}
  {1960}{\natexlab{b}})}\BibitemShut {NoStop}%
\bibitem [{\citenamefont {Caron-Huot}\ \emph {et~al.}(2018)\citenamefont
  {Caron-Huot}, \citenamefont {Dixon}, \citenamefont {von Hippel},
  \citenamefont {McLeod},\ and\ \citenamefont
  {Papathanasiou}}]{Caron-Huot:2018dsv}%
  \BibitemOpen
  \bibfield  {author} {\bibinfo {author} {\bibfnamefont {S.}~\bibnamefont
  {Caron-Huot}}, \bibinfo {author} {\bibfnamefont {L.~J.}\ \bibnamefont
  {Dixon}}, \bibinfo {author} {\bibfnamefont {M.}~\bibnamefont {von Hippel}},
  \bibinfo {author} {\bibfnamefont {A.~J.}\ \bibnamefont {McLeod}},\ and\
  \bibinfo {author} {\bibfnamefont {G.}~\bibnamefont {Papathanasiou}},\
  }\bibfield  {title} {\bibinfo {title} {{The Double Pentaladder Integral to
  All Orders}},\ }\href {https://doi.org/10.1007/JHEP07(2018)170} {\bibfield
  {journal} {\bibinfo  {journal} {JHEP}\ }\textbf {\bibinfo {volume} {07}},\
  \bibinfo {pages} {170}},\ \Eprint {https://arxiv.org/abs/1806.01361}
  {arXiv:1806.01361 [hep-th]} \BibitemShut {NoStop}%
\bibitem [{\citenamefont {He}\ \emph {et~al.}(2022{\natexlab{b}})\citenamefont
  {He}, \citenamefont {Li},\ and\ \citenamefont {Yang}}]{He:2021mme}%
  \BibitemOpen
  \bibfield  {author} {\bibinfo {author} {\bibfnamefont {S.}~\bibnamefont
  {He}}, \bibinfo {author} {\bibfnamefont {Z.}~\bibnamefont {Li}},\ and\
  \bibinfo {author} {\bibfnamefont {Q.}~\bibnamefont {Yang}},\ }\bibfield
  {title} {\bibinfo {title} {{Comments on all-loop constraints for scattering
  amplitudes and Feynman integrals}},\ }\href
  {https://doi.org/10.1007/JHEP01(2022)073} {\bibfield  {journal} {\bibinfo
  {journal} {JHEP}\ }\textbf {\bibinfo {volume} {01}},\ \bibinfo {pages}
  {073}},\ \bibinfo {note} {[Erratum: JHEP 05, 076 (2022)]},\ \Eprint
  {https://arxiv.org/abs/2108.07959} {arXiv:2108.07959 [hep-th]} \BibitemShut
  {NoStop}%
\bibitem [{\citenamefont {Sever}\ \emph
  {et~al.}(2021{\natexlab{a}})\citenamefont {Sever}, \citenamefont {Tumanov},\
  and\ \citenamefont {Wilhelm}}]{Sever:2020jjx}%
  \BibitemOpen
  \bibfield  {author} {\bibinfo {author} {\bibfnamefont {A.}~\bibnamefont
  {Sever}}, \bibinfo {author} {\bibfnamefont {A.~G.}\ \bibnamefont {Tumanov}},\
  and\ \bibinfo {author} {\bibfnamefont {M.}~\bibnamefont {Wilhelm}},\
  }\bibfield  {title} {\bibinfo {title} {{Operator Product Expansion for Form
  Factors}},\ }\href {https://doi.org/10.1103/PhysRevLett.126.031602}
  {\bibfield  {journal} {\bibinfo  {journal} {Phys. Rev. Lett.}\ }\textbf
  {\bibinfo {volume} {126}},\ \bibinfo {pages} {031602} (\bibinfo {year}
  {2021}{\natexlab{a}})},\ \Eprint {https://arxiv.org/abs/2009.11297}
  {arXiv:2009.11297 [hep-th]} \BibitemShut {NoStop}%
\bibitem [{\citenamefont {Sever}\ \emph
  {et~al.}(2021{\natexlab{b}})\citenamefont {Sever}, \citenamefont {Tumanov},\
  and\ \citenamefont {Wilhelm}}]{Sever:2021nsq}%
  \BibitemOpen
  \bibfield  {author} {\bibinfo {author} {\bibfnamefont {A.}~\bibnamefont
  {Sever}}, \bibinfo {author} {\bibfnamefont {A.~G.}\ \bibnamefont {Tumanov}},\
  and\ \bibinfo {author} {\bibfnamefont {M.}~\bibnamefont {Wilhelm}},\
  }\bibfield  {title} {\bibinfo {title} {{An Operator Product Expansion for
  Form Factors II. Born level}},\ }\href
  {https://doi.org/10.1007/JHEP10(2021)071} {\bibfield  {journal} {\bibinfo
  {journal} {JHEP}\ }\textbf {\bibinfo {volume} {10}},\ \bibinfo {pages}
  {071}},\ \Eprint {https://arxiv.org/abs/2105.13367} {arXiv:2105.13367
  [hep-th]} \BibitemShut {NoStop}%
\bibitem [{\citenamefont {Sever}\ \emph {et~al.}(2022)\citenamefont {Sever},
  \citenamefont {Tumanov},\ and\ \citenamefont {Wilhelm}}]{Sever:2021xga}%
  \BibitemOpen
  \bibfield  {author} {\bibinfo {author} {\bibfnamefont {A.}~\bibnamefont
  {Sever}}, \bibinfo {author} {\bibfnamefont {A.~G.}\ \bibnamefont {Tumanov}},\
  and\ \bibinfo {author} {\bibfnamefont {M.}~\bibnamefont {Wilhelm}},\
  }\bibfield  {title} {\bibinfo {title} {{An Operator Product Expansion for
  Form Factors III. Finite Coupling and Multi-Particle Contributions}},\ }\href
  {https://doi.org/10.1007/JHEP03(2022)128} {\bibfield  {journal} {\bibinfo
  {journal} {JHEP}\ }\textbf {\bibinfo {volume} {03}},\ \bibinfo {pages}
  {128}},\ \Eprint {https://arxiv.org/abs/2112.10569} {arXiv:2112.10569
  [hep-th]} \BibitemShut {NoStop}%
\bibitem [{\citenamefont {Basso}\ \emph {et~al.}(2013)\citenamefont {Basso},
  \citenamefont {Sever},\ and\ \citenamefont {Vieira}}]{Basso:2013vsa}%
  \BibitemOpen
  \bibfield  {author} {\bibinfo {author} {\bibfnamefont {B.}~\bibnamefont
  {Basso}}, \bibinfo {author} {\bibfnamefont {A.}~\bibnamefont {Sever}},\ and\
  \bibinfo {author} {\bibfnamefont {P.}~\bibnamefont {Vieira}},\ }\bibfield
  {title} {\bibinfo {title} {{Spacetime and Flux Tube S-Matrices at Finite
  Coupling for $\mathcal{N}=4$ Supersymmetric Yang-Mills Theory}},\ }\href
  {https://doi.org/10.1103/PhysRevLett.111.091602} {\bibfield  {journal}
  {\bibinfo  {journal} {Phys. Rev. Lett.}\ }\textbf {\bibinfo {volume} {111}},\
  \bibinfo {pages} {091602} (\bibinfo {year} {2013})},\ \Eprint
  {https://arxiv.org/abs/1303.1396} {arXiv:1303.1396 [hep-th]} \BibitemShut
  {NoStop}%
\bibitem [{\citenamefont {Alday}\ \emph {et~al.}(2011)\citenamefont {Alday},
  \citenamefont {Gaiotto}, \citenamefont {Maldacena}, \citenamefont {Sever},\
  and\ \citenamefont {Vieira}}]{Alday:2010ku}%
  \BibitemOpen
  \bibfield  {author} {\bibinfo {author} {\bibfnamefont {L.~F.}\ \bibnamefont
  {Alday}}, \bibinfo {author} {\bibfnamefont {D.}~\bibnamefont {Gaiotto}},
  \bibinfo {author} {\bibfnamefont {J.}~\bibnamefont {Maldacena}}, \bibinfo
  {author} {\bibfnamefont {A.}~\bibnamefont {Sever}},\ and\ \bibinfo {author}
  {\bibfnamefont {P.}~\bibnamefont {Vieira}},\ }\bibfield  {title} {\bibinfo
  {title} {{An Operator Product Expansion for Polygonal null Wilson Loops}},\
  }\href {https://doi.org/10.1007/JHEP04(2011)088} {\bibfield  {journal}
  {\bibinfo  {journal} {JHEP}\ }\textbf {\bibinfo {volume} {1104}},\ \bibinfo
  {pages} {088}},\ \Eprint {https://arxiv.org/abs/1006.2788} {arXiv:1006.2788
  [hep-th]} \BibitemShut {NoStop}%
\bibitem [{\citenamefont {Basso}\ \emph
  {et~al.}(2014{\natexlab{a}})\citenamefont {Basso}, \citenamefont {Sever},\
  and\ \citenamefont {Vieira}}]{Basso:2013aha}%
  \BibitemOpen
  \bibfield  {author} {\bibinfo {author} {\bibfnamefont {B.}~\bibnamefont
  {Basso}}, \bibinfo {author} {\bibfnamefont {A.}~\bibnamefont {Sever}},\ and\
  \bibinfo {author} {\bibfnamefont {P.}~\bibnamefont {Vieira}},\ }\bibfield
  {title} {\bibinfo {title} {{Space-time S-matrix and Flux tube S-matrix II.
  Extracting and Matching Data}},\ }\href
  {https://doi.org/10.1007/JHEP01(2014)008} {\bibfield  {journal} {\bibinfo
  {journal} {JHEP}\ }\textbf {\bibinfo {volume} {1401}},\ \bibinfo {pages}
  {008}},\ \Eprint {https://arxiv.org/abs/1306.2058} {arXiv:1306.2058 [hep-th]}
  \BibitemShut {NoStop}%
\bibitem [{\citenamefont {Basso}\ \emph
  {et~al.}(2014{\natexlab{b}})\citenamefont {Basso}, \citenamefont {Sever},\
  and\ \citenamefont {Vieira}}]{Basso:2014koa}%
  \BibitemOpen
  \bibfield  {author} {\bibinfo {author} {\bibfnamefont {B.}~\bibnamefont
  {Basso}}, \bibinfo {author} {\bibfnamefont {A.}~\bibnamefont {Sever}},\ and\
  \bibinfo {author} {\bibfnamefont {P.}~\bibnamefont {Vieira}},\ }\bibfield
  {title} {\bibinfo {title} {{Space-time S-matrix and Flux-tube S-matrix III.
  The two-particle contributions}},\ }\href
  {https://doi.org/10.1007/JHEP08(2014)085} {\bibfield  {journal} {\bibinfo
  {journal} {JHEP}\ }\textbf {\bibinfo {volume} {08}},\ \bibinfo {pages}
  {085}},\ \Eprint {https://arxiv.org/abs/1402.3307} {arXiv:1402.3307 [hep-th]}
  \BibitemShut {NoStop}%
\bibitem [{\citenamefont {Basso}\ \emph
  {et~al.}(2014{\natexlab{c}})\citenamefont {Basso}, \citenamefont {Sever},\
  and\ \citenamefont {Vieira}}]{Basso:2014jfa}%
  \BibitemOpen
  \bibfield  {author} {\bibinfo {author} {\bibfnamefont {B.}~\bibnamefont
  {Basso}}, \bibinfo {author} {\bibfnamefont {A.}~\bibnamefont {Sever}},\ and\
  \bibinfo {author} {\bibfnamefont {P.}~\bibnamefont {Vieira}},\ }\bibfield
  {title} {\bibinfo {title} {{Collinear Limit of Scattering Amplitudes at
  Strong Coupling}},\ }\href {https://doi.org/10.1103/PhysRevLett.113.261604}
  {\bibfield  {journal} {\bibinfo  {journal} {Phys. Rev. Lett.}\ }\textbf
  {\bibinfo {volume} {113}},\ \bibinfo {pages} {261604} (\bibinfo {year}
  {2014}{\natexlab{c}})},\ \Eprint {https://arxiv.org/abs/1405.6350}
  {arXiv:1405.6350 [hep-th]} \BibitemShut {NoStop}%
\bibitem [{\citenamefont {Basso}\ \emph
  {et~al.}(2014{\natexlab{d}})\citenamefont {Basso}, \citenamefont {Sever},\
  and\ \citenamefont {Vieira}}]{Basso:2014nra}%
  \BibitemOpen
  \bibfield  {author} {\bibinfo {author} {\bibfnamefont {B.}~\bibnamefont
  {Basso}}, \bibinfo {author} {\bibfnamefont {A.}~\bibnamefont {Sever}},\ and\
  \bibinfo {author} {\bibfnamefont {P.}~\bibnamefont {Vieira}},\ }\bibfield
  {title} {\bibinfo {title} {{Space-time S-matrix and Flux-tube S-matrix IV.
  Gluons and Fusion}},\ }\href {https://doi.org/10.1007/JHEP09(2014)149}
  {\bibfield  {journal} {\bibinfo  {journal} {JHEP}\ }\textbf {\bibinfo
  {volume} {09}},\ \bibinfo {pages} {149}},\ \Eprint
  {https://arxiv.org/abs/1407.1736} {arXiv:1407.1736 [hep-th]} \BibitemShut
  {NoStop}%
\bibitem [{\citenamefont {Belitsky}(2015{\natexlab{a}})}]{Belitsky:2014sla}%
  \BibitemOpen
  \bibfield  {author} {\bibinfo {author} {\bibfnamefont {A.}~\bibnamefont
  {Belitsky}},\ }\bibfield  {title} {\bibinfo {title} {{Nonsinglet pentagons
  and NMHV amplitudes}},\ }\href
  {https://doi.org/10.1016/j.nuclphysb.2015.05.002} {\bibfield  {journal}
  {\bibinfo  {journal} {Nucl. Phys. B}\ }\textbf {\bibinfo {volume} {896}},\
  \bibinfo {pages} {493} (\bibinfo {year} {2015}{\natexlab{a}})},\ \Eprint
  {https://arxiv.org/abs/1407.2853} {arXiv:1407.2853 [hep-th]} \BibitemShut
  {NoStop}%
\bibitem [{\citenamefont {Belitsky}(2015{\natexlab{b}})}]{Belitsky:2014lta}%
  \BibitemOpen
  \bibfield  {author} {\bibinfo {author} {\bibfnamefont {A.}~\bibnamefont
  {Belitsky}},\ }\bibfield  {title} {\bibinfo {title} {{Fermionic pentagons and
  NMHV hexagon}},\ }\href {https://doi.org/10.1016/j.nuclphysb.2015.02.025}
  {\bibfield  {journal} {\bibinfo  {journal} {Nucl. Phys. B}\ }\textbf
  {\bibinfo {volume} {894}},\ \bibinfo {pages} {108} (\bibinfo {year}
  {2015}{\natexlab{b}})},\ \Eprint {https://arxiv.org/abs/1410.2534}
  {arXiv:1410.2534 [hep-th]} \BibitemShut {NoStop}%
\bibitem [{\citenamefont {Basso}\ \emph
  {et~al.}(2015{\natexlab{a}})\citenamefont {Basso}, \citenamefont {Caetano},
  \citenamefont {Cordova}, \citenamefont {Sever},\ and\ \citenamefont
  {Vieira}}]{Basso:2014hfa}%
  \BibitemOpen
  \bibfield  {author} {\bibinfo {author} {\bibfnamefont {B.}~\bibnamefont
  {Basso}}, \bibinfo {author} {\bibfnamefont {J.}~\bibnamefont {Caetano}},
  \bibinfo {author} {\bibfnamefont {L.}~\bibnamefont {Cordova}}, \bibinfo
  {author} {\bibfnamefont {A.}~\bibnamefont {Sever}},\ and\ \bibinfo {author}
  {\bibfnamefont {P.}~\bibnamefont {Vieira}},\ }\bibfield  {title} {\bibinfo
  {title} {{OPE for all Helicity Amplitudes}},\ }\href
  {https://doi.org/10.1007/JHEP08(2015)018} {\bibfield  {journal} {\bibinfo
  {journal} {JHEP}\ }\textbf {\bibinfo {volume} {08}},\ \bibinfo {pages}
  {018}},\ \Eprint {https://arxiv.org/abs/1412.1132} {arXiv:1412.1132 [hep-th]}
  \BibitemShut {NoStop}%
\bibitem [{\citenamefont {Belitsky}(2015{\natexlab{c}})}]{Belitsky:2015efa}%
  \BibitemOpen
  \bibfield  {author} {\bibinfo {author} {\bibfnamefont {A.~V.}\ \bibnamefont
  {Belitsky}},\ }\bibfield  {title} {\bibinfo {title} {{On factorization of
  multiparticle pentagons}},\ }\href
  {https://doi.org/10.1016/j.nuclphysb.2015.05.024} {\bibfield  {journal}
  {\bibinfo  {journal} {Nucl. Phys.}\ }\textbf {\bibinfo {volume} {B897}},\
  \bibinfo {pages} {346} (\bibinfo {year} {2015}{\natexlab{c}})},\ \Eprint
  {https://arxiv.org/abs/1501.06860} {arXiv:1501.06860 [hep-th]} \BibitemShut
  {NoStop}%
\bibitem [{\citenamefont {Basso}\ \emph
  {et~al.}(2015{\natexlab{b}})\citenamefont {Basso}, \citenamefont {Caetano},
  \citenamefont {Cordova}, \citenamefont {Sever},\ and\ \citenamefont
  {Vieira}}]{Basso:2015rta}%
  \BibitemOpen
  \bibfield  {author} {\bibinfo {author} {\bibfnamefont {B.}~\bibnamefont
  {Basso}}, \bibinfo {author} {\bibfnamefont {J.}~\bibnamefont {Caetano}},
  \bibinfo {author} {\bibfnamefont {L.}~\bibnamefont {Cordova}}, \bibinfo
  {author} {\bibfnamefont {A.}~\bibnamefont {Sever}},\ and\ \bibinfo {author}
  {\bibfnamefont {P.}~\bibnamefont {Vieira}},\ }\bibfield  {title} {\bibinfo
  {title} {{OPE for all Helicity Amplitudes II. Form Factors and Data
  Analysis}},\ }\href {https://doi.org/10.1007/JHEP12(2015)088} {\bibfield
  {journal} {\bibinfo  {journal} {JHEP}\ }\textbf {\bibinfo {volume} {12}},\
  \bibinfo {pages} {088}},\ \Eprint {https://arxiv.org/abs/1508.02987}
  {arXiv:1508.02987 [hep-th]} \BibitemShut {NoStop}%
\bibitem [{\citenamefont {Basso}\ \emph {et~al.}(2016)\citenamefont {Basso},
  \citenamefont {Sever},\ and\ \citenamefont {Vieira}}]{Basso:2015uxa}%
  \BibitemOpen
  \bibfield  {author} {\bibinfo {author} {\bibfnamefont {B.}~\bibnamefont
  {Basso}}, \bibinfo {author} {\bibfnamefont {A.}~\bibnamefont {Sever}},\ and\
  \bibinfo {author} {\bibfnamefont {P.}~\bibnamefont {Vieira}},\ }\bibfield
  {title} {\bibinfo {title} {{Hexagonal Wilson loops in planar ${ \mathcal N
  }=4$ SYM theory at finite coupling}},\ }\href
  {https://doi.org/10.1088/1751-8113/49/41/41LT01} {\bibfield  {journal}
  {\bibinfo  {journal} {J. Phys. A}\ }\textbf {\bibinfo {volume} {49}},\
  \bibinfo {pages} {41LT01} (\bibinfo {year} {2016})},\ \Eprint
  {https://arxiv.org/abs/1508.03045} {arXiv:1508.03045 [hep-th]} \BibitemShut
  {NoStop}%
\bibitem [{\citenamefont {Belitsky}(2017)}]{Belitsky:2016vyq}%
  \BibitemOpen
  \bibfield  {author} {\bibinfo {author} {\bibfnamefont {A.}~\bibnamefont
  {Belitsky}},\ }\bibfield  {title} {\bibinfo {title} {{Matrix pentagons}},\
  }\href {https://doi.org/10.1016/j.nuclphysb.2017.08.011} {\bibfield
  {journal} {\bibinfo  {journal} {Nucl. Phys. B}\ }\textbf {\bibinfo {volume}
  {923}},\ \bibinfo {pages} {588} (\bibinfo {year} {2017})},\ \Eprint
  {https://arxiv.org/abs/1607.06555} {arXiv:1607.06555 [hep-th]} \BibitemShut
  {NoStop}%
\bibitem [{\citenamefont {Goncharov}(2001)}]{Gonch3}%
  \BibitemOpen
  \bibfield  {author} {\bibinfo {author} {\bibfnamefont {A.}~\bibnamefont
  {Goncharov}},\ }\bibfield  {title} {\bibinfo {title} {{Multiple
  polylogarithms and mixed Tate motives}},\ }\href@noop {} {\  (\bibinfo {year}
  {2001})},\ \Eprint {https://arxiv.org/abs/math/0103059} {arXiv:math/0103059
  [math.AG]} \BibitemShut {NoStop}%
\bibitem [{\citenamefont {Brown}(2014)}]{Brown:2013gia}%
  \BibitemOpen
  \bibfield  {author} {\bibinfo {author} {\bibfnamefont {F.}~\bibnamefont
  {Brown}},\ }\bibfield  {title} {\bibinfo {title} {{Single-Valued Motivic
  Periods and Multiple Zeta Values}},\ }\href
  {https://doi.org/10.1017/fms.2014.18} {\bibfield  {journal} {\bibinfo
  {journal} {SIGMA}\ }\textbf {\bibinfo {volume} {2}},\ \bibinfo {pages} {e25}
  (\bibinfo {year} {2014})},\ \Eprint {https://arxiv.org/abs/1309.5309}
  {arXiv:1309.5309 [math.NT]} \BibitemShut {NoStop}%
\bibitem [{\citenamefont {Kotikov}\ and\ \citenamefont
  {Lipatov}(2001)}]{Kotikov:2001sc}%
  \BibitemOpen
  \bibfield  {author} {\bibinfo {author} {\bibfnamefont {A.}~\bibnamefont
  {Kotikov}}\ and\ \bibinfo {author} {\bibfnamefont {L.}~\bibnamefont
  {Lipatov}},\ }\bibfield  {title} {\bibinfo {title} {{DGLAP and BFKL evolution
  equations in the $\mathcal{N}=4$ supersymmetric gauge theory}},\ }in\
  \href@noop {} {\emph {\bibinfo {booktitle} {{35th Annual Winter School on
  Nuclear and Particle Physics}}}}\ (\bibinfo {year} {2001})\ \Eprint
  {https://arxiv.org/abs/hep-ph/0112346} {arXiv:hep-ph/0112346} \BibitemShut
  {NoStop}%
\bibitem [{\citenamefont {Kotikov}\ and\ \citenamefont
  {Lipatov}(2003)}]{Kotikov:2002ab}%
  \BibitemOpen
  \bibfield  {author} {\bibinfo {author} {\bibfnamefont {A.}~\bibnamefont
  {Kotikov}}\ and\ \bibinfo {author} {\bibfnamefont {L.}~\bibnamefont
  {Lipatov}},\ }\bibfield  {title} {\bibinfo {title} {{DGLAP and BFKL equations
  in the $\mathcal{N}=4$ supersymmetric gauge theory}},\ }\href
  {https://doi.org/10.1016/S0550-3213(03)00264-5} {\bibfield  {journal}
  {\bibinfo  {journal} {Nucl. Phys. B}\ }\textbf {\bibinfo {volume} {661}},\
  \bibinfo {pages} {19} (\bibinfo {year} {2003})},\ \bibinfo {note} {[Erratum:
  Nucl.Phys.B 685, 405--407 (2004)]},\ \Eprint
  {https://arxiv.org/abs/hep-ph/0208220} {arXiv:hep-ph/0208220} \BibitemShut
  {NoStop}%
\bibitem [{\citenamefont {Kotikov}\ \emph {et~al.}(2004)\citenamefont
  {Kotikov}, \citenamefont {Lipatov}, \citenamefont {Onishchenko},\ and\
  \citenamefont {Velizhanin}}]{Kotikov:2004er}%
  \BibitemOpen
  \bibfield  {author} {\bibinfo {author} {\bibfnamefont {A.}~\bibnamefont
  {Kotikov}}, \bibinfo {author} {\bibfnamefont {L.}~\bibnamefont {Lipatov}},
  \bibinfo {author} {\bibfnamefont {A.}~\bibnamefont {Onishchenko}},\ and\
  \bibinfo {author} {\bibfnamefont {V.}~\bibnamefont {Velizhanin}},\ }\bibfield
   {title} {\bibinfo {title} {{Three loop universal anomalous dimension of the
  Wilson operators in $\mathcal{N}=4$ SUSY Yang-Mills model}},\ }\href
  {https://doi.org/10.1016/j.physletb.2004.05.078} {\bibfield  {journal}
  {\bibinfo  {journal} {Phys. Lett. B}\ }\textbf {\bibinfo {volume} {595}},\
  \bibinfo {pages} {521} (\bibinfo {year} {2004})},\ \bibinfo {note} {[Erratum:
  Phys.Lett.B 632, 754--756 (2006)]},\ \Eprint
  {https://arxiv.org/abs/hep-th/0404092} {arXiv:hep-th/0404092} \BibitemShut
  {NoStop}%
\bibitem [{\citenamefont {Kotikov}\ \emph {et~al.}(2007)\citenamefont
  {Kotikov}, \citenamefont {Lipatov}, \citenamefont {Rej}, \citenamefont
  {Staudacher},\ and\ \citenamefont {Velizhanin}}]{Kotikov:2007cy}%
  \BibitemOpen
  \bibfield  {author} {\bibinfo {author} {\bibfnamefont {A.}~\bibnamefont
  {Kotikov}}, \bibinfo {author} {\bibfnamefont {L.}~\bibnamefont {Lipatov}},
  \bibinfo {author} {\bibfnamefont {A.}~\bibnamefont {Rej}}, \bibinfo {author}
  {\bibfnamefont {M.}~\bibnamefont {Staudacher}},\ and\ \bibinfo {author}
  {\bibfnamefont {V.}~\bibnamefont {Velizhanin}},\ }\bibfield  {title}
  {\bibinfo {title} {{Dressing and wrapping}},\ }\href
  {https://doi.org/10.1088/1742-5468/2007/10/P10003} {\bibfield  {journal}
  {\bibinfo  {journal} {J. Stat. Mech.}\ }\textbf {\bibinfo {volume} {0710}},\
  \bibinfo {pages} {P10003} (\bibinfo {year} {2007})},\ \Eprint
  {https://arxiv.org/abs/0704.3586} {arXiv:0704.3586 [hep-th]} \BibitemShut
  {NoStop}%
\bibitem [{\citenamefont {Wilczek}(1977)}]{Wilczek:1977zn}%
  \BibitemOpen
  \bibfield  {author} {\bibinfo {author} {\bibfnamefont {F.}~\bibnamefont
  {Wilczek}},\ }\bibfield  {title} {\bibinfo {title} {{Decays of Heavy Vector
  Mesons Into Higgs Particles}},\ }\href
  {https://doi.org/10.1103/PhysRevLett.39.1304} {\bibfield  {journal} {\bibinfo
   {journal} {Phys. Rev. Lett.}\ }\textbf {\bibinfo {volume} {39}},\ \bibinfo
  {pages} {1304} (\bibinfo {year} {1977})}\BibitemShut {NoStop}%
\bibitem [{\citenamefont {Shifman}\ \emph {et~al.}(1978)\citenamefont
  {Shifman}, \citenamefont {Vainshtein},\ and\ \citenamefont
  {Zakharov}}]{Shifman:1978zn}%
  \BibitemOpen
  \bibfield  {author} {\bibinfo {author} {\bibfnamefont {M.~A.}\ \bibnamefont
  {Shifman}}, \bibinfo {author} {\bibfnamefont {A.}~\bibnamefont
  {Vainshtein}},\ and\ \bibinfo {author} {\bibfnamefont {V.~I.}\ \bibnamefont
  {Zakharov}},\ }\bibfield  {title} {\bibinfo {title} {{Remarks on Higgs Boson
  Interactions with Nucleons}},\ }\href
  {https://doi.org/10.1016/0370-2693(78)90481-1} {\bibfield  {journal}
  {\bibinfo  {journal} {Phys. Lett. B}\ }\textbf {\bibinfo {volume} {78}},\
  \bibinfo {pages} {443} (\bibinfo {year} {1978})}\BibitemShut {NoStop}%
\bibitem [{\citenamefont {Dixon}\ \emph {et~al.}(2004)\citenamefont {Dixon},
  \citenamefont {Glover},\ and\ \citenamefont {Khoze}}]{Dixon:2004za}%
  \BibitemOpen
  \bibfield  {author} {\bibinfo {author} {\bibfnamefont {L.~J.}\ \bibnamefont
  {Dixon}}, \bibinfo {author} {\bibfnamefont {E.}~\bibnamefont {Glover}},\ and\
  \bibinfo {author} {\bibfnamefont {V.~V.}\ \bibnamefont {Khoze}},\ }\bibfield
  {title} {\bibinfo {title} {{MHV rules for Higgs plus multi-gluon
  amplitudes}},\ }\href {https://doi.org/10.1088/1126-6708/2004/12/015}
  {\bibfield  {journal} {\bibinfo  {journal} {JHEP}\ }\textbf {\bibinfo
  {volume} {12}},\ \bibinfo {pages} {015}},\ \Eprint
  {https://arxiv.org/abs/hep-th/0411092} {arXiv:hep-th/0411092} \BibitemShut
  {NoStop}%
\bibitem [{\citenamefont {Gehrmann}\ \emph {et~al.}(2012)\citenamefont
  {Gehrmann}, \citenamefont {Jaquier}, \citenamefont {Glover},\ and\
  \citenamefont {Koukoutsakis}}]{Gehrmann:2011aa}%
  \BibitemOpen
  \bibfield  {author} {\bibinfo {author} {\bibfnamefont {T.}~\bibnamefont
  {Gehrmann}}, \bibinfo {author} {\bibfnamefont {M.}~\bibnamefont {Jaquier}},
  \bibinfo {author} {\bibfnamefont {E.}~\bibnamefont {Glover}},\ and\ \bibinfo
  {author} {\bibfnamefont {A.}~\bibnamefont {Koukoutsakis}},\ }\bibfield
  {title} {\bibinfo {title} {{Two-Loop QCD Corrections to the Helicity
  Amplitudes for $H \to$ 3 partons}},\ }\href
  {https://doi.org/10.1007/JHEP02(2012)056} {\bibfield  {journal} {\bibinfo
  {journal} {JHEP}\ }\textbf {\bibinfo {volume} {02}},\ \bibinfo {pages}
  {056}},\ \Eprint {https://arxiv.org/abs/1112.3554} {arXiv:1112.3554 [hep-ph]}
  \BibitemShut {NoStop}%
\bibitem [{\citenamefont {Guo}\ \emph {et~al.}(2022{\natexlab{b}})\citenamefont
  {Guo}, \citenamefont {Jin}, \citenamefont {Wang},\ and\ \citenamefont
  {Yang}}]{Guo:2022pdw}%
  \BibitemOpen
  \bibfield  {author} {\bibinfo {author} {\bibfnamefont {Y.}~\bibnamefont
  {Guo}}, \bibinfo {author} {\bibfnamefont {Q.}~\bibnamefont {Jin}}, \bibinfo
  {author} {\bibfnamefont {L.}~\bibnamefont {Wang}},\ and\ \bibinfo {author}
  {\bibfnamefont {G.}~\bibnamefont {Yang}},\ }\bibfield  {title} {\bibinfo
  {title} {{Deciphering the maximal transcendentality principle via
  bootstrap}},\ }\href {https://doi.org/10.1007/JHEP09(2022)161} {\bibfield
  {journal} {\bibinfo  {journal} {JHEP}\ }\textbf {\bibinfo {volume} {09}},\
  \bibinfo {pages} {161}},\ \Eprint {https://arxiv.org/abs/2205.12969}
  {arXiv:2205.12969 [hep-th]} \BibitemShut {NoStop}%
\bibitem [{\citenamefont {Glover}\ \emph {et~al.}(2008)\citenamefont {Glover},
  \citenamefont {Mastrolia},\ and\ \citenamefont {Williams}}]{Glover:2008ffa}%
  \BibitemOpen
  \bibfield  {author} {\bibinfo {author} {\bibfnamefont {E.~W.~N.}\
  \bibnamefont {Glover}}, \bibinfo {author} {\bibfnamefont {P.}~\bibnamefont
  {Mastrolia}},\ and\ \bibinfo {author} {\bibfnamefont {C.}~\bibnamefont
  {Williams}},\ }\bibfield  {title} {\bibinfo {title} {{One-loop phi-MHV
  amplitudes using the unitarity bootstrap: The General helicity case}},\
  }\href {https://doi.org/10.1088/1126-6708/2008/08/017} {\bibfield  {journal}
  {\bibinfo  {journal} {JHEP}\ }\textbf {\bibinfo {volume} {08}},\ \bibinfo
  {pages} {017}},\ \Eprint {https://arxiv.org/abs/0804.4149} {arXiv:0804.4149
  [hep-ph]} \BibitemShut {NoStop}%
\bibitem [{\citenamefont {Badger}\ \emph {et~al.}(2010)\citenamefont {Badger},
  \citenamefont {Nigel~Glover}, \citenamefont {Mastrolia},\ and\ \citenamefont
  {Williams}}]{Badger:2009hw}%
  \BibitemOpen
  \bibfield  {author} {\bibinfo {author} {\bibfnamefont {S.}~\bibnamefont
  {Badger}}, \bibinfo {author} {\bibfnamefont {E.~W.}\ \bibnamefont
  {Nigel~Glover}}, \bibinfo {author} {\bibfnamefont {P.}~\bibnamefont
  {Mastrolia}},\ and\ \bibinfo {author} {\bibfnamefont {C.}~\bibnamefont
  {Williams}},\ }\bibfield  {title} {\bibinfo {title} {{One-loop Higgs plus
  four gluon amplitudes: Full analytic results}},\ }\href
  {https://doi.org/10.1007/JHEP01(2010)036} {\bibfield  {journal} {\bibinfo
  {journal} {JHEP}\ }\textbf {\bibinfo {volume} {01}},\ \bibinfo {pages}
  {036}},\ \Eprint {https://arxiv.org/abs/0909.4475} {arXiv:0909.4475 [hep-ph]}
  \BibitemShut {NoStop}%
\bibitem [{\citenamefont {Badger}\ \emph {et~al.}(2007)\citenamefont {Badger},
  \citenamefont {Glover},\ and\ \citenamefont {Risager}}]{Badger:2007si}%
  \BibitemOpen
  \bibfield  {author} {\bibinfo {author} {\bibfnamefont {S.~D.}\ \bibnamefont
  {Badger}}, \bibinfo {author} {\bibfnamefont {E.~W.~N.}\ \bibnamefont
  {Glover}},\ and\ \bibinfo {author} {\bibfnamefont {K.}~\bibnamefont
  {Risager}},\ }\bibfield  {title} {\bibinfo {title} {{One-loop phi-MHV
  amplitudes using the unitarity bootstrap}},\ }\href
  {https://doi.org/10.1088/1126-6708/2007/07/066} {\bibfield  {journal}
  {\bibinfo  {journal} {JHEP}\ }\textbf {\bibinfo {volume} {07}},\ \bibinfo
  {pages} {066}},\ \Eprint {https://arxiv.org/abs/0704.3914} {arXiv:0704.3914
  [hep-ph]} \BibitemShut {NoStop}%
\bibitem [{\citenamefont {Badger}\ and\ \citenamefont
  {Glover}(2006)}]{Badger:2006us}%
  \BibitemOpen
  \bibfield  {author} {\bibinfo {author} {\bibfnamefont {S.~D.}\ \bibnamefont
  {Badger}}\ and\ \bibinfo {author} {\bibfnamefont {E.~W.~N.}\ \bibnamefont
  {Glover}},\ }\bibfield  {title} {\bibinfo {title} {{One-loop helicity
  amplitudes for H $\rightarrow$ gluons: The All-minus configuration}},\ }\href
  {https://doi.org/10.1016/j.nuclphysbps.2006.09.030} {\bibfield  {journal}
  {\bibinfo  {journal} {Nucl. Phys. B Proc. Suppl.}\ }\textbf {\bibinfo
  {volume} {160}},\ \bibinfo {pages} {71} (\bibinfo {year} {2006})},\ \Eprint
  {https://arxiv.org/abs/hep-ph/0607139} {arXiv:hep-ph/0607139} \BibitemShut
  {NoStop}%
\bibitem [{\citenamefont {Bern}\ \emph {et~al.}(2004)\citenamefont {Bern},
  \citenamefont {Dixon},\ and\ \citenamefont {Kosower}}]{Bern:2004cz}%
  \BibitemOpen
  \bibfield  {author} {\bibinfo {author} {\bibfnamefont {Z.}~\bibnamefont
  {Bern}}, \bibinfo {author} {\bibfnamefont {L.~J.}\ \bibnamefont {Dixon}},\
  and\ \bibinfo {author} {\bibfnamefont {D.~A.}\ \bibnamefont {Kosower}},\
  }\bibfield  {title} {\bibinfo {title} {{Two-loop $g \rightarrow gg$ splitting
  amplitudes in QCD}},\ }\href {https://doi.org/10.1088/1126-6708/2004/08/012}
  {\bibfield  {journal} {\bibinfo  {journal} {JHEP}\ }\textbf {\bibinfo
  {volume} {08}},\ \bibinfo {pages} {012}},\ \Eprint
  {https://arxiv.org/abs/hep-ph/0404293} {arXiv:hep-ph/0404293} \BibitemShut
  {NoStop}%
\bibitem [{\citenamefont {Abreu}\ \emph {et~al.}(2014)\citenamefont {Abreu},
  \citenamefont {Britto}, \citenamefont {Duhr},\ and\ \citenamefont
  {Gardi}}]{Abreu:2014cla}%
  \BibitemOpen
  \bibfield  {author} {\bibinfo {author} {\bibfnamefont {S.}~\bibnamefont
  {Abreu}}, \bibinfo {author} {\bibfnamefont {R.}~\bibnamefont {Britto}},
  \bibinfo {author} {\bibfnamefont {C.}~\bibnamefont {Duhr}},\ and\ \bibinfo
  {author} {\bibfnamefont {E.}~\bibnamefont {Gardi}},\ }\bibfield  {title}
  {\bibinfo {title} {{From multiple unitarity cuts to the coproduct of Feynman
  integrals}},\ }\href {https://doi.org/10.1007/JHEP10(2014)125} {\bibfield
  {journal} {\bibinfo  {journal} {JHEP}\ }\textbf {\bibinfo {volume} {10}},\
  \bibinfo {pages} {125}},\ \Eprint {https://arxiv.org/abs/1401.3546}
  {arXiv:1401.3546 [hep-th]} \BibitemShut {NoStop}%
\bibitem [{\citenamefont {Bloch}\ and\ \citenamefont
  {Kreimer}(2015)}]{Bloch:2015efx}%
  \BibitemOpen
  \bibfield  {author} {\bibinfo {author} {\bibfnamefont {S.}~\bibnamefont
  {Bloch}}\ and\ \bibinfo {author} {\bibfnamefont {D.}~\bibnamefont
  {Kreimer}},\ }\bibfield  {title} {\bibinfo {title} {{Cutkosky Rules and Outer
  Space}},\ }\href@noop {} {\  (\bibinfo {year} {2015})},\ \Eprint
  {https://arxiv.org/abs/1512.01705} {arXiv:1512.01705 [hep-th]} \BibitemShut
  {NoStop}%
\bibitem [{\citenamefont {Abreu}\ \emph {et~al.}(2017)\citenamefont {Abreu},
  \citenamefont {Britto}, \citenamefont {Duhr},\ and\ \citenamefont
  {Gardi}}]{Abreu:2017enx}%
  \BibitemOpen
  \bibfield  {author} {\bibinfo {author} {\bibfnamefont {S.}~\bibnamefont
  {Abreu}}, \bibinfo {author} {\bibfnamefont {R.}~\bibnamefont {Britto}},
  \bibinfo {author} {\bibfnamefont {C.}~\bibnamefont {Duhr}},\ and\ \bibinfo
  {author} {\bibfnamefont {E.}~\bibnamefont {Gardi}},\ }\bibfield  {title}
  {\bibinfo {title} {{Algebraic Structure of Cut Feynman Integrals and the
  Diagrammatic Coaction}},\ }\href
  {https://doi.org/10.1103/PhysRevLett.119.051601} {\bibfield  {journal}
  {\bibinfo  {journal} {Phys. Rev. Lett.}\ }\textbf {\bibinfo {volume} {119}},\
  \bibinfo {pages} {051601} (\bibinfo {year} {2017})},\ \Eprint
  {https://arxiv.org/abs/1703.05064} {arXiv:1703.05064 [hep-th]} \BibitemShut
  {NoStop}%
\bibitem [{\citenamefont {Bourjaily}\ \emph {et~al.}(2021)\citenamefont
  {Bourjaily}, \citenamefont {Hannesdottir}, \citenamefont {McLeod},
  \citenamefont {Schwartz},\ and\ \citenamefont {Vergu}}]{Bourjaily:2020wvq}%
  \BibitemOpen
  \bibfield  {author} {\bibinfo {author} {\bibfnamefont {J.~L.}\ \bibnamefont
  {Bourjaily}}, \bibinfo {author} {\bibfnamefont {H.}~\bibnamefont
  {Hannesdottir}}, \bibinfo {author} {\bibfnamefont {A.~J.}\ \bibnamefont
  {McLeod}}, \bibinfo {author} {\bibfnamefont {M.~D.}\ \bibnamefont
  {Schwartz}},\ and\ \bibinfo {author} {\bibfnamefont {C.}~\bibnamefont
  {Vergu}},\ }\bibfield  {title} {\bibinfo {title} {{Sequential Discontinuities
  of Feynman Integrals and the Monodromy Group}},\ }\href
  {https://doi.org/10.1007/JHEP01(2021)205} {\bibfield  {journal} {\bibinfo
  {journal} {JHEP}\ }\textbf {\bibinfo {volume} {01}},\ \bibinfo {pages}
  {205}},\ \Eprint {https://arxiv.org/abs/2007.13747} {arXiv:2007.13747
  [hep-th]} \BibitemShut {NoStop}%
\bibitem [{\citenamefont {Hannesdottir}\ \emph
  {et~al.}(2022{\natexlab{a}})\citenamefont {Hannesdottir}, \citenamefont
  {McLeod}, \citenamefont {Schwartz},\ and\ \citenamefont
  {Vergu}}]{Hannesdottir:2021kpd}%
  \BibitemOpen
  \bibfield  {author} {\bibinfo {author} {\bibfnamefont {H.~S.}\ \bibnamefont
  {Hannesdottir}}, \bibinfo {author} {\bibfnamefont {A.~J.}\ \bibnamefont
  {McLeod}}, \bibinfo {author} {\bibfnamefont {M.~D.}\ \bibnamefont
  {Schwartz}},\ and\ \bibinfo {author} {\bibfnamefont {C.}~\bibnamefont
  {Vergu}},\ }\bibfield  {title} {\bibinfo {title} {{Implications of the Landau
  equations for iterated integrals}},\ }\href
  {https://doi.org/10.1103/PhysRevD.105.L061701} {\bibfield  {journal}
  {\bibinfo  {journal} {Phys. Rev. D}\ }\textbf {\bibinfo {volume} {105}},\
  \bibinfo {pages} {L061701} (\bibinfo {year} {2022}{\natexlab{a}})},\ \Eprint
  {https://arxiv.org/abs/2109.09744} {arXiv:2109.09744 [hep-th]} \BibitemShut
  {NoStop}%
\bibitem [{\citenamefont {Hannesdottir}\ and\ \citenamefont
  {Mizera}(2023)}]{Hannesdottir:2022bmo}%
  \BibitemOpen
  \bibfield  {author} {\bibinfo {author} {\bibfnamefont {H.~S.}\ \bibnamefont
  {Hannesdottir}}\ and\ \bibinfo {author} {\bibfnamefont {S.}~\bibnamefont
  {Mizera}},\ }\href {https://doi.org/10.1007/978-3-031-18258-7} {\emph
  {\bibinfo {title} {{What is the i\ensuremath{\varepsilon} for the
  S-matrix?}}}},\ SpringerBriefs in Physics\ (\bibinfo  {publisher}
  {Springer},\ \bibinfo {year} {2023})\ \Eprint
  {https://arxiv.org/abs/2204.02988} {arXiv:2204.02988 [hep-th]} \BibitemShut
  {NoStop}%
\bibitem [{\citenamefont
  {M\"uhlbauer}(2022{\natexlab{a}})}]{Muhlbauer:2022ylo}%
  \BibitemOpen
  \bibfield  {author} {\bibinfo {author} {\bibfnamefont {M.}~\bibnamefont
  {M\"uhlbauer}},\ }\bibfield  {title} {\bibinfo {title}
  {{Cutkosky\textquoteright{}s theorem for massive one-loop Feynman integrals:
  part 1}},\ }\href {https://doi.org/10.1007/s11005-022-01612-4} {\bibfield
  {journal} {\bibinfo  {journal} {Lett. Math. Phys.}\ }\textbf {\bibinfo
  {volume} {112}},\ \bibinfo {pages} {118} (\bibinfo {year}
  {2022}{\natexlab{a}})},\ \Eprint {https://arxiv.org/abs/2206.08402}
  {arXiv:2206.08402 [math-ph]} \BibitemShut {NoStop}%
\bibitem [{\citenamefont {Hannesdottir}\ \emph
  {et~al.}(2022{\natexlab{b}})\citenamefont {Hannesdottir}, \citenamefont
  {McLeod}, \citenamefont {Schwartz},\ and\ \citenamefont
  {Vergu}}]{Hannesdottir:2022xki}%
  \BibitemOpen
  \bibfield  {author} {\bibinfo {author} {\bibfnamefont {H.~S.}\ \bibnamefont
  {Hannesdottir}}, \bibinfo {author} {\bibfnamefont {A.~J.}\ \bibnamefont
  {McLeod}}, \bibinfo {author} {\bibfnamefont {M.~D.}\ \bibnamefont
  {Schwartz}},\ and\ \bibinfo {author} {\bibfnamefont {C.}~\bibnamefont
  {Vergu}},\ }\bibfield  {title} {\bibinfo {title} {{Constraints on Sequential
  Discontinuities from the Geometry of On-shell Spaces}},\ }\href@noop {} {\
  (\bibinfo {year} {2022}{\natexlab{b}})},\ \Eprint
  {https://arxiv.org/abs/2211.07633} {arXiv:2211.07633 [hep-th]} \BibitemShut
  {NoStop}%
\bibitem [{\citenamefont {Bourjaily}\ \emph {et~al.}(2022)\citenamefont
  {Bourjaily}, \citenamefont {Vergu},\ and\ \citenamefont {von
  Hippel}}]{Bourjaily:2022vti}%
  \BibitemOpen
  \bibfield  {author} {\bibinfo {author} {\bibfnamefont {J.~L.}\ \bibnamefont
  {Bourjaily}}, \bibinfo {author} {\bibfnamefont {C.}~\bibnamefont {Vergu}},\
  and\ \bibinfo {author} {\bibfnamefont {M.}~\bibnamefont {von Hippel}},\
  }\bibfield  {title} {\bibinfo {title} {{Landau Singularities and Higher-Order
  Roots}},\ }\href@noop {} {\  (\bibinfo {year} {2022})},\ \Eprint
  {https://arxiv.org/abs/2208.12765} {arXiv:2208.12765 [hep-th]} \BibitemShut
  {NoStop}%
\bibitem [{\citenamefont
  {M\"uhlbauer}(2022{\natexlab{b}})}]{Muhlbauer:2022zzz}%
  \BibitemOpen
  \bibfield  {author} {\bibinfo {author} {\bibfnamefont {M.}~\bibnamefont
  {M\"uhlbauer}},\ }\bibfield  {title} {\bibinfo {title} {{On the Homology of
  Unions of Certain Non-Degenerate Quadrics in General Position}},\ }\href@noop
  {} {\  (\bibinfo {year} {2022}{\natexlab{b}})},\ \Eprint
  {https://arxiv.org/abs/2211.06683} {arXiv:2211.06683 [math-ph]} \BibitemShut
  {NoStop}%
\bibitem [{\citenamefont {Binosi}\ \emph {et~al.}(2009)\citenamefont {Binosi},
  \citenamefont {Collins}, \citenamefont {Kaufhold},\ and\ \citenamefont
  {Theussl}}]{Binosi:2008ig}%
  \BibitemOpen
  \bibfield  {author} {\bibinfo {author} {\bibfnamefont {D.}~\bibnamefont
  {Binosi}}, \bibinfo {author} {\bibfnamefont {J.}~\bibnamefont {Collins}},
  \bibinfo {author} {\bibfnamefont {C.}~\bibnamefont {Kaufhold}},\ and\
  \bibinfo {author} {\bibfnamefont {L.}~\bibnamefont {Theussl}},\ }\bibfield
  {title} {\bibinfo {title} {{JaxoDraw: A Graphical user interface for drawing
  Feynman diagrams. Version 2.0 release notes}},\ }\href
  {https://doi.org/10.1016/j.cpc.2009.02.020} {\bibfield  {journal} {\bibinfo
  {journal} {Comput. Phys. Commun.}\ }\textbf {\bibinfo {volume} {180}},\
  \bibinfo {pages} {1709} (\bibinfo {year} {2009})},\ \Eprint
  {https://arxiv.org/abs/0811.4113} {arXiv:0811.4113 [hep-ph]} \BibitemShut
  {NoStop}%
\end{thebibliography}

%

\end{document}